\documentstyle[psfig,eqsecnum,floats,aps]{revtex}

\def\spi{i^o}
\def\ut#1{\rlap{\lower1ex\hbox{$\sim$}}{#1}}
\def\l{\ell}
\def\N{{\cal N}}

\def\bF{{\bf F}}
\def\bE{{\bf E}}
\def\bA{{\bf A}}
\def\M{{\bf M}}

\def\pb#1{\rlap{\lower1.5ex\hbox{$\longleftarrow$}}{#1}}
\def\dpb#1{\rlap{\lower1.5ex\hbox{$\Longleftarrow$}}{#1}}
\def\spb#1{\rlap{\lower1.0ex\hbox{$\leftarrow$}}{#1}}
\def\sdpb#1{\rlap{\lower1.0ex\hbox{$\Leftarrow$}}{#1}}
\def\d{{\rm d}}

\def\bar{\overline}
\def\ba{\begin{eqnarray}}
\def\ea{\end{eqnarray}}
\def\be{\begin{equation}}
\def\ee{\end{equation}}
\def\Lab{{\cal L}}
\def\={\mathrel{\widehat\mathalpha{=}}}
\def\puto#1{\rlap{\raise.5ex\hbox{\char'27}}{#1}}
\newcommand{\teta}{\rlap{\lower2ex\hbox{$\,\tilde{}$}}\eta{}}
\newcommand{\tieta}{\tilde{\eta}}

\preprint{\vbox{\baselineskip=12pt
\rightline{gr-qc/0002078}
\rightline{ICN-UNAM/00-04}
}}

\begin{document}
\draft
\title{
Einstein-Yang-Mills Isolated Horizons:\\
Phase Space, Mechanics, Hair and Conjectures}
\author {
Alejandro\ Corichi\thanks{E-mail:
corichi@nuclecu.unam.mx}, Ulises
Nucamendi and Daniel Sudarsky\thanks{E-mail:
sudarsky@nuclecu.unam.mx}
}

\address{Instituto de Ciencias Nucleares\\
Universidad Nacional Aut\'onoma de M\'exico\\
A. Postal 70-543, M\'exico D.F. 04510, M\'exico.}

\maketitle

\begin{abstract}
The concept of ``Isolated Horizon" has been recently used to provide
a  full Hamiltonian treatment of black holes.
It has been applied successfully to
the cases of {\it non-rotating}, {\it non-distorted}  black holes in
Einstein Vacuum, Einstein-Maxwell and Einstein-Maxwell-Dilaton Theories.
In this note, it is investigated
the extent to which the framework can be generalized
to the case of
non-Abelian gauge theories where `hairy black holes' are known to exist.
It is found that this extension is indeed possible, despite the fact
that in general, there is no
`canonical normalization' yielding a preferred  Horizon Mass.
In particular the zeroth and first laws are
established for all normalizations. Colored static spherically
symmetric black hole solutions to
the Einstein-Yang-Mills equations are considered from
this perspective. A canonical formula
for the Horizon Mass of such black holes is found.
This analysis is used to obtain
nontrivial relations between the masses of the colored black holes
and the regular solitonic solutions in Einstein-Yang-Mills
theory. A general
testing bed for the instability of hairy black holes in general
non-linear theories is suggested. As an example,
the embedded Abelian magnetic solutions are considered. It is
shown that, within this framework, the total energy is also
positive and thus, the solutions are potentially unstable.
Finally, it is discussed which elements would be needed to place
the Isolated Horizons framework for Einstein-Yang-Mills theory
in the same footing as the previously analyzed cases.
Motivated by these considerations
 and using the fact that the Isolated Horizons framework
seems to be the appropriate language to state  uniqueness
and completeness
conjectures for the EYM equations --in terms of the
horizon charges--, two such conjectures are put forward.
\end{abstract}

\section{Introduction}
\label{sec1}

The nonperturbative quantum geometry program also known as
``loop quantum gravity",
has met recently with substantial success in obtaining a
calculation of the statistical mechanical entropy of a
non-rotating black hole that
accounts for  its phenomenological identification
with  $\frac{1}{4} A$ \cite{abck}.
In so doing it was necessary to introduce a complete classical
Hamiltonian treatment of Black Holes.
This was accomplished by generalizing and
properly defining the sector of the theory that is going
to be treated. This work was guided by the need
to start with a well defined action that would be differentiable
in the sector under consideration. This lead those
 authors to the specialization
of the notion of Trapping Horizons \cite{sh} of Hayward's to that of
`Isolated Horizons'. Physically the idea is to represent
``horizons in internal equilibrium and decoupled from what is outside".

The zeroth and first laws of black hole mechanics refer to
equilibrium situations and small departures therefrom.  Therefore, in
the standard treatments \cite{1,2,3,4,chru,mh} one restricts oneself to
stationary space-times admitting event horizons and perturbations off
such space-times. The isolated
horizons (IH) framework, which is tailored to more
general physical situations was introduced in \cite{ack} and the
corresponding zeroth and first laws of black hole mechanics were
established.
\cite{abf1,abf2,acdil}.
This framework generalizes the treatment of  black hole mechanics in
two directions.  First, the notion of event horizons is replaced by
that of `isolated horizons' which can be defined
quasi-locally, unlike the former which can only be defined
retroactively, after having access to the entire space-time history.
Second, the underlying
space-time need not admit any Killing field; isolated horizons need
not be Killing horizons.  The static event horizons normally used in
black hole mechanics \cite{1,2,3}  are  special cases of isolated
horizons.  Moreover, because one can now admit gravitational and matter
radiation, there are many more examples.  In particular, while the
manifold ${\cal S}$ of static space-times admitting event horizons in the
Einstein-Maxwell theory is finite dimensional, the manifold ${\cal IH}$
of space-times admitting isolated horizons is \textit{infinite}
dimensional \cite{abf2}.

The resulting formalism is then, not only a step in the
construction of the quantum theory for the sector but also
a new tool for studying classical aspects of black holes.
When restricted to the Static sector of the theory,
it leads, for example, to an improvement in the treatment of the
physical process version of the first law
of black hole mechanics \cite{abf1,abf2}.
The formalism has so far only been applied to
the Einstein-Maxwell system (with and without dilaton field), of which
the known exact solutions (the Reissner
Nordstrom solutions in EM and the
 so called Gibbons-Maeda solutions in EMD)
are particular examples.

In this work we will  explore the extent  to which  this formalism can
be extended to
the Einstein-Yang-Mills theory where, in the static
spherically symmetric sector, one finds
the so called Colored Black Hole solutions \cite{gv,kunzle,bizon}.
The motivation is many-fold. First, we are interested in
studying the robustness of  the isolated horizons formalism
in its ability  to treat other theories, specially those where
`hairy' solutions are know to exist.
{}From the isolated horizons perspective, this poses a special
challenge, since now there is an apparent
tension given by the  mismatch between the number of
conserved charges at infinity and at the horizon. In the colored
black hole solutions, the only non-zero
`charge' at infinity is the ADM
mass, but the gauge field is non-vanishing at the horizon
and it contributes to the `horizon magnetic charge'. Is the
isolated horizon framework robust enough to deal with
this situation and to resolve this tension?
Second, we would like to know whether the
general formalism allows us to learn new facts
about the Static Spherically Symmetric (SSS) solutions.
In particular, we
want to understand  what differences, if any, appear when we
treat static but unstable  black hole solutions,
which is the case for the Abelian magnetic
solutions and the higher $n$ colored black holes.
Finally, we want to investigate  whether the formalism can
--as can be expected given the fact that
it yields a satisfactory treatment of the inner boundary in
Hamiltonian terms-- allow us to take the limit when the horizon
area goes to zero and connect
the black hole solutions with the regular, solitonic,
solutions that are known to also exist in this theory.

We will in fact find  answers to all of these questions and
puzzles. We find however, some subtleties that need to be stressed.
In particular, the fact that there are no Spherically
Symmetric solutions for some values of the horizon parameters, poses
a challenge to the isolated horizon framework, since there seems to
be no canonical value for the Horizon
Mass of the black hole in this situation.
Thus, if the EYM system is to be  in the same `status' as
Einstein-Maxwell system, some yet unknown part of the current scenario
would have to yield to the tension mentioned above.
The most natural  resolution  would entail the validity of
a `uniqueness conjecture' for Static solutions
and a `completeness conjecture' for the existence of Stationary solutions,
that we will put forward.
Their validity would guarantee the complete consistency of
the formalism.

On the other hand, we will show  some new results
regarding SSS solutions as a  nontrivial `application' of the
formalism  as it allows us to predict a relation between the ADM mass of
a static black hole solution, its Horizon mass and the ADM mass of the
corresponding solitonic solutions. These relations
can be corroborated by numerical computations
(consistent with the reported results in the literature).
Therefore,  these `coincidences' can be viewed in a sense,
as a check on the formalism.

This paper is organized as follows. In Section~\ref{sec2},
we recall some basic facts about the EYM system and about
the known SSS solutions, both
Abelian and non-Abelian. In Sec.~\ref{sec3}
we specify the isolated horizon boundary conditions
that we impose, making the necessary adjustments to
incorporate non-Abelian gauge fields. Section~\ref{sec4}
deals with the action principle of the theory in the presence
of the horizon as an inner boundary, and the specification of
the phase space of the theory. In Section~\ref{sec5} we consider
the definition of surface gravity in the absence of a Killing
field and we show the zeroth law of BH mechanics for general
isolated horizons. The definition of
horizon mass together with the first law are studied  in
Section~\ref{sec6}. Section~\ref{sec7} is devoted to the
completeness conjecture and some of its implication for
stationary solutions.
In Section~\ref{sec8},
we return to the study of Static Spherically
Symmetric solution to the EYM equations from the perspective of
isolated black holes, and show some new results.
Finally, we end with a discussion in Section~\ref{sec9}.

Throughout the paper, we use units in which $c=G=1$, and the
abstract index notation of Penrose's \cite{wald}, except in
Sec.~\ref{sec6} where we use differential forms.

\section{Einstein-Yang-Mills and Static Black Holes}
\label{sec2}

This section has two parts. In the first one, we recall some basic
facts about the Einstein-Yang-Mills system. In the second part,
we briefly review the Static Spherically Symmetric solutions
to the EYM equations focusing on black hole solutions.

\subsection{Einstein-Yang-Mills System}

In the Einstein-Yang-Mills system, the gravitational
part of the action, $S_{\rm Grav}$ is given by:
\be S_{\rm Grav} =-\frac{1}{16\pi}\int_{{\M}}\sqrt{-g}
\;R\, \d^4x\, ,
\ee
and the matter part of the action  is given by:
\be S_{\rm YM} (\bA) =-\frac{1}{16\pi g_{\rm YM}^2}\int_{{\M}}\sqrt{-g}
\;[\bF^i_{ab}\bF_i^{ab}]\,\d^4x\label{dil:act}\, ,
\ee
where the abstract indices $a,b,\ldots$ denote space-time objects and the
indices $i,j,\ldots$ are internal indices in the Lie algebra of the gauge
group $G$. In this paper we shall consider $G=SU(2)$.
The field strength $\bF_{ab}$ is given by
$\bF^i_{ab}=2\nabla_{[a}\bA^i_{b]}+
{\epsilon^i}_{jk}\bA_a^j\bA_b^k$, that is, the
curvature of the Lie algebra valued one form $\bA_a^i$.
The total action  $S_{\rm Tot}$ is given by
\be S_{\rm Tot}=S_{\rm Grav} + S_{\rm YM}. \ee
We remind the reader that in the case of a Non-Abelian Yang Mills
theory there
is dimension-full parameter $g_{\rm YM}$ that, unlike
the Abelian  case,  can not be absorbed in the ``gauge" fields.
This endows the
full theory with a natural scale at the classical level
i.e. ${g_{\rm YM}}^2$ has dimensions of mass.

The equations of motion that follow from  $S_{\rm Tot}$ are:
\ba
D_a {}\bF^{iab} = 0,
\label{dil:eom1}\\
R_{ab}= 2\left(\bF^i_{ac}{\bF_{ib}}^c -\frac{1}{4}
g_{ab}\bF^2  \right),
\label{dil:eom3}
\ea
where $\bF^2=\bF^i_{ab}\bF_i^{ab}$, and $D_a$ is the generalized
covariant derivative defined by $\bA$. Furthermore we have the
 Bianchi identity for the Yang Mills sector:
\ba
D_{[c}\bF^i_{ab]}=0,
\ea
The dual field tensor is
given by ${}^*\bF^i_{ab}=\frac{1}{2}{\epsilon_{ab}}^{cd}\bF^i_{cd}$,
where $\epsilon_{abcd}$ is the canonical volume element associated with
$g_{ab}$.

Note that in contrast with the Einstein-Maxwell system, where,
under appropriate fall-of conditions, the
conserved electric and magnetic charges can be defined at infinity,
the naive expressions,
\be
\hat{Q}^i:=\frac{1}{4\pi}\oint_{S_\infty}{}^*\bF^i, \qquad
\hat{P}^i:=\frac{1}{4\pi}\oint_{S_\infty}\bF^i \label{QyP}\, ,
\ee
fail to be gauge invariant. It is only in the presence of a globally
defined isometry that the conserved charges might be invariantly defined
(i.e. a natural gauge might be chosen).
We can nevertheless define new gauge invariant quantities, for any
two-sphere
$S$ as follows,
\be
Q_S:=\frac{1}{4\pi}\oint_{S}|{}^*\bF|, \qquad
P_S:=\frac{1}{4\pi}\oint_{S}|\bF| \label{QoP}\, ,
\ee
where $|\bF|_{ab}$ is the two form defined in the following way:
 we take $\epsilon_{ab} $ the area two form associated with the 2-sphere $S$
and define $f^i = \bF^i_{ab} \epsilon^{ab}$.
Then $|\bF|_{ab} =\sqrt{\sum (f^i)^2} \epsilon_{ab}$.
$|{}^*\bF|_{ab}$ is analogously defined. In what follows, we shall refer
to $Q_S$ and $P_S$ as the electric and magnetic charge contained `within' 
$S$ respectively.

\subsection{Static Solutions}

Static Spherically Symmetric (SSS) solutions to the EYM equations
representing black hole space-times are known to exist
in different situations (for  a recent review see \cite{review}
and references therein).

A standard parameterization for the metric and gauge potential
is given by,
\ba
\label{SSS}
\d s^2 &=& -\left(1-\frac{2m(r)}{r}\right)e^{-2\delta(r)}\d t^2
+\left(1-\frac{2m(r)}{r}\right)^{-1}\d r^2+ r^2\d\Omega^2\, ,
\label{param1}\\
\bA &=& a\tau_3\d t +b\tau_3\d r+(w\tau_1+d\tau_2)\d \theta
+(\cot\theta\tau_3+w\tau_2-d\tau_1)\sin\theta\,\d\phi\, ,
\label{param2}
\ea
where $a,b,w$ and $d$ are functions of $(r,t)$.

One can then look for either static regular solutions by requiring
$m(r)<r/2$ for all $r\ge 0$ \cite{BK} or for static black hole solutions
with horizon at $r=r_H $ by requiring $m(r_H) = r_H/2$ and
$m(r)<r/2$ for all $r\ge r_H$\cite{kunzle,bizon}. Solutions of both
types are found in the purely magnetic sector, for which,
with a further gauge choice one can set $a=b=d=0$,
and $w$ becomes a function of $r$ only.

The regular or solitonic solutions compose a discrete set
parameterized by the
number of nodes of the function $w(r)$ and
are characterized by their ADM mass whose scale is set by
${g_{\rm YM}}^2$ \cite{bizon2}.

On the other hand, for every value of $r_H$, one finds also
two classes of black hole solutions: Abelian and
non-Abelian.

\subsubsection{Abelian Solutions}

The first class of solutions  is given by what are essentially
Abelian solutions embedded in SU(2).
Within the Abelian sector we have either
electrically charged or magnetically charged
solutions. Unlike Einstein-Maxwell theory where
the well known Maxwell duality exists, in EYM
there is no such duality and one is not allowed to
treat them on `equal footing'.

The electrically charged solutions with
electric charge $Q$ are given by
the standard solutions that can be described by choosing $a\not =0$
and $b=d=w=0$ in (\ref{param2}). The YM potential is of the
form $\bA=\frac{Q}{r}\tau_3\,\d t$ and the metric is given by
(\ref{param1}) with the functions  $m(r)=M-({Q}^2)/2r$ and $\delta=0$.
Note that these represent a two parameter family of `RN solutions'
with parameters $M$ and $Q$.

The magnetically charged solutions with
magnetic charge $P$ are precisely the
Reissner-Nordstrom solutions
given by $w=0$,  $m(r)=M-({P}^2)/2r$ and $\delta=0$. Since
we are considering the Abelian case,
the magnetic charge $P$ is not arbitrary but can take only
one value $P=1$. Note that
the YM field strength,
in the `magnetic' sector of the  EYM theory takes the form,
\be
\bF=w^\prime\tau_1\d r\wedge \d\theta+
w^\prime\tau_2\sin\theta\; \d r\wedge \d\phi -(1-w^2)\tau_3\sin\theta\;
\d\theta\wedge\d\phi\, .
\label{2.11}
\ee
Thus, for the RN solutions where $w=0$, we get from (\ref{2.11}) that
$P=1$ for any sphere containing the black hole.
The magnetically charged solutions are then parameterized by only
one charge, namely, the ADM mass $M$.

These two sectors share the $Q=P=0$ solution corresponding to
the Schwarzschild solution. One can also construct dyonic solutions with
both electric charge $Q$ and unit magnetic charge. In all these solutions,
one has to satisfy the inequality $r^2_H\geq Q^2 + P^2$, in order to
have black hole solutions with no naked singularities.

\subsubsection{Non-Abelian Colored Black Holes}

These solutions correspond to the
purely magnetic case, where, for each value of the horizon
area the equations  have a discrete
number of solutions which are strictly  Non-Abelian in nature
 (i.e. do not exit in the Abelian regime). These are
 labeled
by an integer $n$ that represents the number of nodes of the function
$w(r)$.
The lowest mode, $n=0$,
represents the Schwarzschild solution. Therefore,
the solution can be completely parameterized by two numbers
$(a_\Delta,n)$, the horizon area and the integer $n$.
All this solutions, for $n>0$ are unstable under perturbations
\cite{uns}.

On the other hand it is known that there are no
nontrivial dyonic solutions
(i.e solutions with electric and magnetic fields)
in the spherically symmetric sector\cite{Popp}.

{}From a historical perspective, these were the first examples of `hairy
black holes' \cite{gv,kunzle,bizon}. This is because the electric
and magnetic charges
(\ref{QyP}) are both zero, so the only parameter at infinity
is the ADM mass. If the no-hair conjecture were valid for
the EYM system, the
specification of $M_{\rm ADM}$ would suffice to characterize the
solution completely. However, this is not the case, since for a
given value of the ADM mass, there exist a countable number of
{\it different} solutions, labeled by $n$.
Equivalently we can label these solutions by the ADM
mass $M_{\rm ADM}$ and $n$.

Even when the charges at infinity are not enough to specify static black
holes uniquely, one might still hope to
have quasi-local quantities defined at
the horizon, that are in a sense, good coordinates for the
manifold ${\cal S}$ of static
solutions.  Indeed, we shall put forward in the following sections
a `quasi-local  uniqueness conjecture' ($C1$):
{\it All static BH solutions are characterized by its
horizon parameters arising from the `isolated horizon'
framework}. Let us refer to these quantities
defined at the horizon as `quasi-local parameters'.
In theories where no hair is present, as is the case of the
Einstein-Maxwell-Dilaton system, the number of `quasi-local
parameters' equals the number of parameters at infinity labeling
the static solutions \cite{acdil}.
Thus, stating a uniqueness conjecture in this
theory is insensitive as to whether one is postulating it in terms
of quantities at infinity (the standard viewpoint), or in terms
of `quasi-local charges'. Our proposal is that, for general
theories, one should state the  postulate in
terms of purely quasi-local quantities.
In the EYM system, the quasi-local charges are $a_\Delta$, the horizon
area, $Q_\Delta$ and $P_\Delta$, the horizon electric and magnetic charges
respectively. In this case the first conjecture $C1$ reads:
Given a triple of parameters $(a_\Delta,Q_\Delta,P_\Delta)$ for
which a SSS solutions exists, then the solution is unique.
However, note an apparent tension in this suggestion:
Given the mismatch of the number of parameters at the horizon and
at infinity, there is room for inconsistency when formulating,
say, the laws of thermodynamics. This is because the number of
`independent' parameters is different, when one
considers charges at infinity or quasi-local parameters for, say,
SSS colored solutions.
As we shall see in  Section~\ref{sec6}, the
nature of the problem can be made precise
within the isolated horizons framework, and some ideas can be put
forward to understand the origin of the difficulty.

For the convenience of the reader, in the next section we shall
recall the notion of isolated
horizons as defined in \cite{ack,abf2,acdil} and explore some of
its consequences for EYM system of interest to this paper.

\section{Boundary Conditions and Consequences}
\label{sec3}

Let us recall the notion of isolated horizons $\Delta$ in
general, and include in its definition the relevant
modifications to incorporate
the Einstein-Yang-Mills system.
The basic boundary conditions defining $\Delta$
are the same as those introduced in \cite{acdil}.

Let us begin by recalling some notation. Fix any null surface
${\N}$, topologically $S^2\times R$, and consider foliations of $\N$
by families of spatial 2-spheres.  Given a
foliation, we parameterize its leaves by $v={\rm const}$ such that $v$
increases to the future and set $n_a = -\nabla_a v$.  Under a
reparametrization $v\mapsto F(v)$, we have $n_a\mapsto F'(v) n_a$ with
$F'(v) >0$.  Thus, every foliation comes equipped with an equivalence
class $[n_a]$ of normals $n_a$ related by rescalings which are constant on
each leaf.%
\footnote{These 1-form fields $n_a$ are defined intrinsically on $\N$.
We can extend each $n_a$ to the full space-time uniquely by demanding
that the extended 1-form be null. However, in this paper, we will not
need this extension.}
Also recall that, given any one $n_a$, we can uniquely select a vector
field $\l^a$
which is normal to $\N$ and satisfies $\l^a n_a = -1$. (Thus, $\l^a$ is
future-pointing.)  If we change the parameterization, $\l^a$ transforms
via: $\l^a \mapsto (F'(v))^{-1} \l^a$.  Thus, given a foliation, we
acquire an equivalence class $[\l^a, n_a]$ of pairs, $(\l^a, n_a)$,
of vector fields and 1-forms on $\N$ subject to the relation $(\l^a,
n_a) \sim (G^{-1} \l^a, G n_a)$, where $G$ is any positive function on
$\N$ which is constant on each leaf of the foliation.  Given a pair
$(\l^a, n_a)$ in the equivalence class, we introduce a complex vector
field $m^a$ on $\N$, tangential to each leaf in the foliation, such that
$m^a \bar{m}_a = 1$.  (By construction, $m^a \l_a = m^a n_a = 0$ on
$\N$.)  The vector field $m^a$ is unique up to a phase factor.  With
this structure at hand, we now look at the main Definition.

\bigskip\noindent {\it Definition:} The internal boundary $\Delta$ of a
space-time $(\M, g_{ab})$ will be said to represent {\it a
non-rotating isolated horizon} provided the following conditions hold%
\footnote{Throughout this paper, the symbol
{${\mathrel{\widehat\mathalpha{=}}}$} will denote equality at points
of $\Delta$.  For fields defined throughout space-time, an under-arrow
will denote pull-back to $\Delta$. The part of the Newman-Penrose
framework \cite{pr} used in this paper is summarized in the Appendices
A and B of \cite{abf2}.}:
\begin{itemize}
\item{(i)} {\it Manifold conditions:} $\Delta$ is a null surface,
topologically $S^{2}\times R$.
\item{(ii)} {\it Dynamical conditions:} All field equations
hold at $\Delta$.
\item{(iii)} {\it Main conditions:} $\Delta$ admits a foliation such that
the Newman-Penrose coefficients associated with the corresponding direction
fields $[\l^a,n_a]$ on $\Delta$ satisfy the following conditions:\\
(iii.a) $\rho \= -\bar{m}^a {m}^b \nabla_a\l_b$, the expansion of $[\l^a]$,
vanishes on $\Delta$.\\
(iii.b) $\lambda \= \bar{m}^a \bar{m}^b \nabla_a n_b$ and $\pi \= \l^a
\bar{m}^b \nabla_a n_b$ vanish on $\Delta$ and the expansion $\mu :=
m^a\bar{m}^b \nabla_a n_b$ of $n_a$ is negative
\footnote{For simplicity, in this paper we focus on black-hole-type
horizons rather than cosmological ones. To incorporate interesting
cosmological horizons, one has to weaken this condition and allow
the possibility that $\mu$ is everywhere positive on $\Delta$. See
\cite{abf2}.}
and constant on each leaf of the foliation.
\item{(iv)} {\it Conditions on matter:} The Yang-Mills
field $\bF$ is such that
\be
|{\rm Re}\,\phi_1|, \quad{\rm and}\quad |{\rm Im}\,\phi_1|,
\ee
are constant on each leaf of the foliation introduced in condition (iii).
(Recall that $\phi^i_1 \= \frac{1}{2}\,
m^a\bar{m}^b (\bF - i\, {}^\star\!\bF)^i_{ab}$)
where $|{\rm Re}\,\phi_1|:=\sqrt{\sum_i (\rm Re \phi^i_1)
(\rm Re \phi^i_1)}$, and $|{\rm Im}\,\phi_1|$ is analogously defined.
\end{itemize}

The first two conditions are quite tame: (i) simply asks that $\Delta$
be null and have appropriate topology while (ii) is completely
analogous to the dynamical condition imposed at infinity.  As the
terminology suggests, (iii.a) and (iii.b) are the most important
conditions.  Note first that, if a pair $(\l^a,n_a)$ in the equivalence
class $[\l^a,n_a]$ associated with the foliation satisfies these
conditions, so does any other pair, $((G(v))^{-1} \l^a ,\, G(V) n_a)$.
Thus, the conditions are well-defined.  They are motivated by the
following considerations.  Condition (iii.a) captures the idea that
the horizon is isolated without having to refer to a Killing field.
In particular, it implies that the area of each 2-sphere leaf in the
foliation be the same.  We will denote this area by $a_\Delta$ and
define the \textit{horizon radius} $r_\Delta$ via $a_\Delta = 4\pi
r^2_{\Delta}$.

Condition (iii.b) has three sets of implications.  First, one can show
that if, as required, one can find a foliation of $\Delta$ satisfying
(iii.b), that foliation is \textit{unique}.  (In the SSS family, as
one might expect, this condition selects the foliation to which the
rotational Killing fields are tangential.)  Second, it implies
that the imaginary part of (the Newman-Penrose Weyl component)
$\Psi_2$, which captures angular momentum, vanishes and thus restricts
us to {\it non-rotating} horizons. Third, the requirement that the
expansion $\mu$ of $n^a$ be negative implies that $\Delta$ is a
\textit{future} horizon rather than \textit{past}
horizon \cite{sh}. Finally, consider the
spherical symmetry requirement on the Yang-Mills field component
$\phi^i_1$ given by condition (iv).
While this condition is a strong restriction, it can be
motivated by analogy with the Einstein-Maxwell case.
(For further motivation
and remarks on these conditions, see \cite{ack,abf2}.)

Since these conditions are \textit{local} to $\Delta$,
the notion of an isolated horizon is quasi-local; in
particular, one does not need an entire space-time history to locate
an isolated horizon.  Furthermore, the boundary conditions allow for
the presence of radiation in the exterior region, thus, space-times
admitting isolated horizons need not admit any Killing field
\cite{jl}.  Indeed, the manifold ${\cal IH}$ of solutions to field
equations
admitting isolated horizons is infinite dimensional
\cite{abf2}).

In spite of this generality, boundary conditions place
strong restrictions on the structure of various fields \textit{at}
$\Delta$.  Let us begin with conditions on the Yang-Mills
field.  The stress-energy tensor $T_{ab}$ of $\bF$ satisfies
the dominant energy condition.  Hence, on $\Delta$, $-T_{ab}\l^b$ is a
future directed, causal vector field.  Now, using the Raychaudhuri
equation and field equations \textit{at} $\Delta$ (condition (ii) of
the Definition), we conclude $T_{ab}\l^a\l^b \=0$.  By expanding out
this expression (see Eq (\ref{dil:eom3})) we obtain
\be \label{3.2}
\bF^i_{ab} \= \phi^i_1 2 (\l_{[a}\, n_{b]} - m_{[a}\, \bar{m}_{b]}) +
\phi^i_2 2 (m_{[a}\, \l_{b]})+{\rm CC}\, ,
\ee
for \textit{some} complex algebra-valued functions
 $\phi^i_1$ and $\phi^i_2$ (the only non-vanishing
Newman-Penrose components of $\bF^i_{ab}$) on $\Delta$, where ${\rm CC}$
stands for `the complex conjugate term'.  These equations say that
there is no flux of Yang-Mills radiation
across $\Delta$.   Finally, condition (iv) in the Definition implies
\be |{\rm Re}\,\phi_1| \=  \frac{2\pi}{a_\Delta}\,
{Q}_\Delta , \qquad
|{\rm Im}\,\phi_1| \=  \frac{2\pi}{a_\Delta}\,
{P}_\Delta,\ee
where $Q_\Delta$ is the electric charge and $P_\Delta$ the magnetic
charge at the horizon as defined by
(\ref{QoP}).
Thus the
boundary conditions severely restrict the form of matter fields at
$\Delta$.
 The component $\phi^i_0 = -\l^a m^b \bF^i_{ab}$ of the YM field vanishes
and the gauge invariant components
of $\phi^i_1$ are completely determined by the
electric and magnetic charges.  However, the component $\phi^i_2$ of the
YM field is unconstrained.

Restrictions imposed on space-time curvature at $\Delta$ are
essentially the same as in Ref \cite{abf2}.%
\footnote{This is because these restrictions were obtained assuming
rather general conditions on the matter stress-energy which are satisfied
in EYM.  The derivation of some of these results involve long
calculations and a topological result on the Chern-class of the $SO(2)$
connection associated with the dyad $(m,\bar{m})$. See \cite{abf2}.}
Results relevant to this paper can be summarized as follows.  In the
Newman-Penrose notation, for the Ricci tensor components, we have:
\ba \label{ricci}
\Phi_{00} &=& \frac{1}{2}\, R_{ab} \l^a\l^b \= 0, \quad
\Phi_{01} = \frac{1}{2}\, R_{ab}\l^a m^b \= 0,
\nonumber\\
\Phi_{11} &=& \frac{1}{4}R_{ab} (\l^an^b +m^a\bar{m}^b) \=
 8\pi^2 \,
\frac{(Q_\Delta^2+P^2_\Delta)}{a_\Delta^2}, \quad R \= 0 ,
\ea
where $R$ is the scalar curvature. The Weyl tensor components satisfy
\ba\label{weyl}
\Psi_0 &=& C_{abcd}\l^am^b\l^cm^d \= 0, \qquad
\Psi_1 = C_{abcd}\l^am^b\l^cn^d \= 0\, ,\nonumber\\
\Psi_2 &=& C_{abcd}\l^a m^b\bar{m}^c n^d \= \Phi_{11}  -
\frac{2\pi}{a_\Delta}\, .
\ea
Furthermore,
\be
\Psi_3 \= \Phi_{21}, \quad{\hbox {\rm that is}}\quad
C_{abcd}\l^an^b\bar{m}^cn^d \= \frac{1}{2} R_{ab}\bar{m}^an^b.
\ee
As expected, for the SSS solutions discussed in
Section \ref{sec2}, these conditions are
satisfied. We note that even when the curvature components $\Psi_2$
and $\Phi_{11}$ are the only ones different from
zero in SSS solutions, for a general isolated horizon
other curvature components
(like $\Psi_3$ and $\Psi_4$) may be `dynamical', i.e., vary along the
integral curves of $\l$.

In the Einstein-Maxwell-Dilaton system
\cite{acdil}, we have a
set of parameters $(a_\Delta,Q_\Delta,P_\Delta,\phi_\Delta)$ at the
horizon (and the same number at infinity), and
those parameters were naturally selected as the horizon parameters.
In the EYM case
we have seen that we can define electric and
magnetic charges at the horizon $(Q_\Delta, P_\Delta)$. The boundary
conditions ensure that this quantities are constant along the horizon
$\Delta$ (and explicitly gauge invariant). Thus, it is natural to
use the triplet $a_\Delta, Q_\Delta, P_\Delta$ to parameterize general
EYM isolated horizons. Let us denote by $\Lab$ the space of
horizon parameters with coordinates $(a_\Delta, Q_\Delta, P_\Delta)$,
with the following restrictions: $a_\Delta > 4\pi
(Q_\Delta^2+P_\Delta^2)$ and $Q_\Delta, P_\Delta\in
[0,\infty)$.

\section{Action and Phase Space}
\label{sec4}

The gravitational action has been shown to be differentiable, with
respect to variations respecting the isolated horizons boundary
conditions, in \cite{ack,abf2}. We refer the reader to those papers
for details. One important property of the variational principle is that
one is varying, in the pure gravitational case, histories with a
fixed value $a^o_\Delta$ of the horizon area.
We also need to ensure that the matter action is
differentiable and work out the Hamiltonian framework for the matter
sector as well. This could require imposition of additional boundary
conditions on matter fields, but
as we will see, the conditions already imposed on
matter,  described in Sec.~\ref{sec3} will be enough.
The purpose of this section is concentrate on the variational
principle for
the Yang-Mills field and work out the Hamiltonian description.

Since we require that field equations hold on $\Delta$, the
gravitational boundary conditions already imply certain restrictions
on the behavior of YM fields there. As
noted in Section~\ref{sec3}, boundary conditions imply that several
components of the Ricci tensor vanish on $\Delta$,
and that the curvature
tensor $\bF$ has the form (\ref{3.2}).
In particular $ |{\rm Re}(\phi_1)|$
and $|{\rm Im}(\phi_1)|$ are spherically
symmetric on the preferred cross-sections.  In that case,
$|{\bf \phi}_1|$ can be
expressed in terms of the electric and magnetic charges,
$Q_\Delta$ and $P_\Delta$ of the isolated horizon,
\begin{eqnarray}\label{el-charge}
Q :\= -{1\over 4\pi} \oint_{S_v} |{}^*{\bf F}| \=
{1\over 2\pi} \oint_S\,
|{\rm Re}\,{\bf \phi}_1| \,\, {}^{2}\epsilon\, , \\
\label{magn-charge}
P :\= -{1\over 4\pi} \oint_{S_v} |{\bf F}| \= {1\over 2\pi}
\oint_{S_v} |{\rm Im}\, {\bf \phi}_1| \,\,
{}^{2}\epsilon\, ,
\end{eqnarray}
(Here $S_v$ are the 2-spheres $v={\rm const}$ in the preferred
foliation. The minus sign in front of the first integrals in
(\ref{el-charge}) and (\ref{magn-charge}) arise because we have oriented
$S_v$ such that the radial normal is in-going rather than outgoing.)

Let's now  discuss the action principle.
For the same reasons that the area $a_\Delta$ is
kept fixed in the variational principle, we will now
restrict ourselves to histories for which the values of electric and
magnetic charges on the horizon are fixed to $Q^o_{\Delta}$
and $P^o_{\Delta}$
respectively. To make the action principle well-defined, we need to
impose suitable boundary conditions on the YM fields. Conditions
at infinity are the standard ones given in \cite{sud:wal}. To find
boundary conditions on $\Delta$, let us consider the standard
YM bulk action:
\begin{equation}
S_{\rm YM} = -{1\over 16\pi} \int_{\M} \sqrt{-g} \;
{\bf F}^i_{ab} {\bf F}_i^{ab} \d^4x \, .
\end{equation}
Variation of $S_{\rm YM}$ yields
\begin{equation}\label{ch:variation}
\delta\left(S_{\rm YM}\right)
%{\bf F}^i\wedge {}^*{\bf F}_i\right)
= - {1\over 4\pi} \int_{\M}   (D_a
{\bf F}^{abi})\delta {\bf A}_i\; \sqrt{-g}\,\d^4x +
{1\over 4\pi} \int_{\partial \M}
\delta {\bf A}^i_{[a}\, {}^*{\bf F}_{bc]}^i \,\tieta^{abc}\d^3x.
\end{equation}
As usual, the bulk term provides the equations of motion provided the
surface term vanishes.
The boundary term (\ref{ch:variation})
at infinity is the usual one and is dealt with in the standard manner
\cite{sud:wal}. When evaluated
at the horizon, the boundary term (\ref{ch:variation}) does not
automatically vanish. On $\Delta$, the
boundary term in (\ref{ch:variation}) can be written
as,
\be
\frac{3}{4\pi}\int_{\Delta}\delta \bA^i_a{}^*\bF_{ibc}
\l^{[c}\tilde{\eta}^{ab]}\;\d v\,\d^2x\, .
\ee
 Now, (\ref{3.2}) implies that on $\Delta$ the
pull-back of $\bF^i_{ab}\l^b$  vanishes.
Then, the horizon contribution  reduces to
\begin{equation}\label{ch:7}
\int \d v\oint_{S_v}\, \delta(\bA_{ia} \l^a)\, {}^*\bF_{ab}^i\tieta^{ab}
\d^2x\, ,
\end{equation}
where, as before, $v$ is the affine parameter
(with respect to the horizon metric) along the integral
curves of $\l^a$ such that $v={\rm const}$ define the preferred
foliation of $\Delta$ and $S_v$ are the 2-spheres in this foliation.
Now, since isolated horizons are to be thought of as
``non-dynamical'', it is natural to ask that the gauge field $\bA^i_a$
be invariant under the action of $\l$. This is equivalent to ask that
${\cal L}_\l \spb{{\bf A}^i_a} \= \spb{D_a{\cal V}^i_\l}$ 
be satisfied \cite{b:s}, 
for ${\cal V}^i_\l$ the gauge generator.  This condition is
naturally satisfied since the pull-back of $\bF^i_{ab}\l^b$
to the horizon $\Delta$  vanishes. 
Furthermore, the form of the boundary term
(\ref{ch:7}) suggests that we fix gauge so that
$({\bf A}^i\cdot \l):=A^i_a\l^a$ is
proportional to ${\rm Re}(\phi^i_1)$, that is,
$({\bf A}^i\cdot \l)=c\, {\rm Re}(\phi^i_1)$
for $c$ a constant on our space of histories
\footnote{This gauge condition was independently found by
Ashtekar, Fairhurst and Krishnan
in the more general context of \textit{distorted} horizons\cite{afk}.
We thank S. Fairhurst for communicating their results prior to
publication.}.
The norm of $({\bf A}^i\cdot \l)$ as we shall see in following sections,
is determined by the consistency of
the Hamiltonian formalism (and will take
its standard value in the static solutions, when they exist).
Then the boundary term (\ref{ch:7})
arising in the variation of the action
vanishes, i.e., the bulk action itself is differentiable and the
action principle is well-defined. Note that the permissible gauge
transformations are now restricted: If $\bA_a \mapsto \bA_a + D_a f$, the
generating `function' $f^i$ has to  satisfy
$l^a D_a f^i \= 0$ on $\Delta$ (as well as satisfy standard
fall-off conditions at infinity).

\begin{figure}
\centerline{
\hbox{\psfig{figure=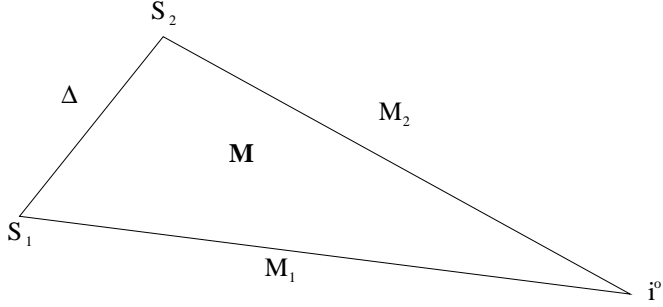,height=4cm}}}
\bigskip
\caption{Region $\M$ of space-time considered in the
variational principle is bounded by two partial Cauchy surfaces $M_1$
and $M_2$.  They intersect the isolated horizon $\Delta$ in preferred
2-spheres $S_1$ and $S_2$ and extend to spatial infinity
$\spi$.}\label{exam}
\end{figure}

Let us now provide a summary of the structure of the phase space of
Yang-Mills fields.  Fix a foliation of ${\M}$ by a family of
space-like 3-surfaces $M_t$ (level surfaces of a time function $t$)
which intersect $\Delta$ in the preferred 2-spheres.
Fix a normalization for the vector field $\l^a$.  Denote by $t^a$
a ``time-evolution'' vector field which is not tangential to the
foliation with affine parameter $t$ which tends to a unit
time-translation at infinity and to the vector field $l^a$ on
$\Delta$. The canonical conjugate moment is given by,
\be
\tilde{{\bf \Pi}}_i^a=\frac{\sqrt{h}}{4\pi}h^{ac}{\bf n}^b\bF_{ibc},
\ee
where $h_{ab}$ is the intrinsic metric on $M_t$ induced from $g_{ab}$,
and ${\bf n}^a$ is the normal to $M_t$.
In terms of the lapse and shift fields $N$ and $N^a$ defined
by $t^a$, the Legendre transform of the action yields:
\ba \label{lt2}
S_{\rm EM}&=&\frac{1}{4\pi}\int \d t \int_{M_t}\,\d^3x\left(
\tilde{{\bf \Pi}}_i^a\nabla_a(\bA^i\cdot\l)-N^dF^i_{ad}\tilde{{\bf
\Pi}}^{ia}
+\right.\\ \nonumber
&{}&\left.
\frac{N}{2\sqrt{h}}\;h_{ab}\tilde{{\bf \Pi}}^{ia}\tilde{{\bf \Pi}}_i^b+
\frac{N}{2}\sqrt{h}\;F^{ab}_iF^i_{ab}\right)\, ,
\ea
where the 2-form $\bE_{iab}$ is the pull-back
 to $M$ of ${}^{*}\bF_{iab}$,
${}^{(*)}{\bE}^i_a:=\frac{1}{2}{\epsilon_a}^{bc}{\bE}^i_{bc}$, and
${}^{(*)}{\bF}^i_a:=\frac{1}{2}{\epsilon_a}^{bc}{\bF}^i_{bc}$,
with $\epsilon_{abc}$ the volume form defined by $h_{ab}$.
The two-form $\bE^i_{ab}$ is related to the canonically conjugate
moment as follows: $\bE^i_{ab}=\frac{1}{4\pi}\teta_{abc}
\tilde{{\bf \Pi}}^{ic}$.

Thus, as usual, the phase space consists of pairs $(\bA_a^i, \bE_{abi})$
on the 3-manifold $M_t$, subject
to boundary conditions, where the connection $\bA^i_a$
is now the pull-back to $M_t$ of the Yang-Mills 4-potential and the
2-form $\bE^i_{ab}$ is the dual of the  electric field vector
density. These fields are subject to boundary conditions. On any
horizon 2-sphere $S_v$, conditions (iv)
must hold ensuring that the pull-backs of $|\bF|_{ab}$ and $|\bE|_{ab}$ are
spherically symmetric%
\footnote{at the horizon $\Delta$, the natural
parameter $v$ and the `time' parameter $t$ coincide.}.
(Since $(\bA^i\cdot \l)$ appears as a Legendre
multiplier in (\ref{lt2}) it is not part of  the
phase space variables at $\Delta$.)  At infinity, $\bA_a^i, \bE_{iab}$ are
subject to the appropriate boundary conditions
that ensure asymptotic flatness
of the metric and are general enough to include the nontrivial
EYM solutions mentioned in section~\ref{sec3}.
The conditions for the YM sector are \cite{sud:wal}:
\be
A_a^i =A_a^i(\theta,\phi)/r +o(r^{-1}), \qquad
 F_{ab}^i =F_{ab}^i(\theta,\phi)/r +o(r^{-2})\, .
\label{ConYMifty}
\ee
The symplectic structure on this YM-sector of phase
space can be read off from (\ref{lt2}):
\begin{equation}
\Omega|_{({\bf A},{\bf E})}\left(\delta_1, \delta_2 \right)
= {1\over 4\pi} \int_M
\left[ \delta_1 \tilde{\bf E}^a_i\;\delta_2 {\bf A}^i_a -
\delta_2 \tilde{\bf E}_i^a \; \delta_1 {\bf A}^i_a \right].
\end{equation}
(The asymptotic conditions ensure that the integrals converge.)
As usual, there is one first class constraint, $D_{[a}\bE^i_{bc]}=0$, which
generates gauge transformations: Under the canonical transformation
generated by $\int f_i\, D_{[a}{\bf E}^i_{bc]}\tieta^{abc}\d^3x$,
the canonical fields transform, as
usual, via $\bA^i_a \mapsto \bA^i_a + D_af^i$ and ${\bf E}_{iab}$ `rotates'.
Note that our boundary conditions allow the generating function $f_i$ to
be non-trivial on the (intersection of $M_t$ with) $\Delta$; the
smeared constraint function is still differentiable. Thus, as
in the gravitational case, the gauge degrees of freedom do {\it not}
become physical in this framework.

Let us summarize. By imposing a set of boundary conditions on the
horizon, to be satisfied by {\it all} histories in the variational
principle, we arrived at the phase space of EYM isolated horizons
${\cal IH}_0$. This phase space can be seen as the (gauge equivalence
class of) solutions to the equations of motion with constant and fixed
quasi-local parameters $a^o_{\Delta}$, $Q^o_{\Delta}$ and $P^o_{\Delta}$.
For the purpose of the variational principle
(in the sense that is well defined and yields the correct
equation of motion), it is enough to
consider configurations with fixed values of the horizon parameters.
However, as we shall see in following sections, when considering the
Hamiltonian formulation --essential for the formulation of the
first law-- one needs to extend the space of isolated horizons from
${\cal IH}_0$ to ${\cal IH}$, where {\it all} possible values of the
quasi-local parameters are considered \cite{abf2}.

\section{Surface Gravity and the Zeroth Law}
\label{sec5}

In each SSS solution there is a unique time-translational Killing
field $t^a$ which is unit at infinity.  As usual, surface gravity
$\kappa_{\rm SSS}$ is defined in terms of its acceleration at the
horizon: $t^a \nabla_a t^b\= \kappa_{\rm SSS}\, t^b$.
Unlike the Einstein-Maxwell-Dilaton theory where the knowledge
of the exact solutions allow us to write
$\kappa$ in terms of the parameters of the solutions, for the
YM field we do not have a closed form for $\kappa$. The only
expression we have at our disposal is the general formula found
by Visser for SSS space-times of the form (\ref{SSS}),
independently of the matter content
of the theory, given by the expression  \cite{visser},
\be
\kappa_{\rm SSS}=\frac{1}{2r_{\rm h}}e^{-\delta(r_{\rm h})}\left[
1-2m^\prime(r_{\rm h})\right]\, .
\label{ksss}
\ee
{}From the perspective of the isolated horizon
framework, $\kappa$ is the acceleration of the properly normalized
null normal $\l^a$ to $\Delta$ \cite{abf1,abf2}. In the SSS solutions,
$\Delta$ happens to be a Killing horizon and we can select a unique
vector field $\l^a$ from the the equivalence class $[\l^a]$ simply by
setting $\l^a \= t^a$. Then $\kappa_{\rm SSS}$ is the acceleration of
this specific $\l^a$.  In the case of general isolated horizons, the
challenge is to find a prescription to single out a preferred $\l^a$,
without reference to any Killing field.  The strategy we
would like to adopt has two
steps and is motivated by the one used for isolated
horizons in dilaton gravity \cite{acdil}. However, as we shall see
below, we will encounter some difficulties which make the
EYM case more subtle than the systems previously studied.

In the first step, we will normalize $\l^a$ {\it only} up to a constant,
leaving a rescaling freedom $\l^a \mapsto \l^{\prime a} = c\l^a$,
where $c$ is
a constant on $\Delta$ but may depend on the parameters $r_\Delta,
{Q}_\Delta, P_\Delta$ of the isolated horizon.  For each such
$\l^a$, we can define the surface gravity $\kappa_{\l}$ {\it relative to
that} $\l^a$ via $\l^a\nabla_a \l^b \= \kappa_{\l} \l^b$.  Rescaling of
$\l^a$ now induces to a `gauge transformation' in $\kappa$: $\kappa_{\l}
\mapsto \kappa_{\l^\prime} = c \kappa_{\l}$.  (Recall that in the
general Newman-Penrose framework, $\kappa$ is a connection component
and therefore undergoes the standard gauge transformations under a
change of the null tetrad.  By fixing $\l^a$ up to a constant rescaling,
we have reduced the general gauge freedom to that of a constant
rescaling.)  Since the zeroth law only says that the surface gravity
is constant on $\Delta$, if it holds for one $\l^a$, it holds for every
$\l^{\prime a} = c\l^a$.  Thus, for the zeroth law, it is in fact
\textit{not} essential to get rid of the rescaling freedom.

Recall that the isolated horizon is naturally equipped with
equivalence classes $[\l,n]$ of vector and co-vector fields, subject
to the relation: $(\l, n) \sim (G^{-1}\l, G n)$ for any positive
function $G \equiv G(v)$ on $\Delta$. As we already mentioned, our
first task is to reduce the
freedom in the choice of $G(v)$ to that of a constant.  We use the
same strategy as in \cite{abf1,abf2}.  (For motivation, see
\cite{abf2}.)  Recall that $\mu$, the expansion of $n$ is strictly
negative  and constant on each leaf of the preferred
foliation; $\mu \equiv \mu(v) <0$.  It is easy to verify that
\be n^a \mapsto G(v) n^a \,\,{\rm implies}\quad \mu \mapsto G(v)
\mu(v).  \ee
Hence, we can \textit{always} use the $G(v)$ freedom to set $\mu \=
{\rm const}$.  This condition restricts the family of $(\l^a,n_a)$ pairs
and reduces the equivalence relation to $(\l^a,n_a) \sim (c\l^a, c^{-1}n_a)$
where $c$ is any constant on $\Delta$. We will denote the restricted
equivalence class by $[\l^a,n_a]_R$. In the second step, we can
\textit{arbitrarily fix} the numerical value of $\mu$
in terms of the parameters
of the isolated horizon and eliminate the rescaling freedom
altogether, thereby selecting,
\textit{for each choice} of $\mu$,
a canonical pair $(\l^a,n_a)$ on each isolated horizon.

With the equivalence class $[\l^a,n_a]_R$ at our disposal, as discussed
above, we can define a surface gravity $\kappa_{\l}$ via $\l^a
\nabla_a \l^b \= \kappa_{\l} \l^b$.  Constancy of $\kappa_{\l}$ on
$\Delta$ follows from the same arguments that were used in
\cite{abf2}.  For completeness, let us briefly recall the structure of
that proof.  First, using conditions on derivatives of $l^a,n_a$
introduced in the Definition, one can express the self-dual part of
the Riemann curvature in terms of $\kappa_{\l}, \nabla_a\kappa_{\l}, \mu$
(and another field which is not relevant to this discussion).
Comparing this expression to the standard Newman-Penrose expansion of
the self-dual curvature tensor in terms of curvature scalars \cite{pr},
and using the fact that certain curvature scalars vanish on $\Delta$
(see (\ref{ricci}) and (\ref{weyl})), one can conclude
\be \label{4.13}
(\nabla_{[a}\kappa_{\l}) n_{b]} \= 0, \quad{\rm and}\quad
\kappa_{\l} \= \frac{\Psi_2}{\mu}.
\ee
The first equation implies that $\kappa_{\l}$ is spherically symmetric.
Hence, it only remains to show that ${\cal L}_{\l}\,\kappa_{\l} \=0$.
Since $\mu$ is now a constant on $\Delta$, it suffices to show that
${\cal L}_\l \Psi_2  =0$. Now, the (second) Bianchi identity implies that
\be \label{bi}
{\cal L}_{\l}\, \left(\Psi_2 - \Phi_{11} \right) \= 0\, .
\ee
Finally, using (\ref{3.2}), we conclude: $\Phi_{11} = 8\pi^2 \,
 \frac{(Q^2_\Delta+P_\Delta^2)}{a_\Delta^2}$.
Thus, $\Phi_{11}$ is
constants on $\Delta$.  Combining these results, we conclude ${\cal
L}_{\l} \kappa_{\l} \= 0$, whence $\kappa_{\l}$ is constant on
$\Delta$.  This establishes the zeroth law.

Let us now consider the second step
in fixing the normalization of $\l$.  So far, we have only required
that $\mu$ be a (negative) constant but not fixed its value.  Under
the rescaling $\mu \mapsto c^{-1}\mu$ we have: $\l \mapsto c\l$, and
$\kappa_{\l} \mapsto c \kappa_{\l}$.  Hence, the remaining rescaling
freedom in $\l$ and $\kappa$ can be exhausted simply by fixing the
value of $\mu$ in terms of the isolated horizon parameters.
We would like to single out a canonical choice, and the
obvious strategy is to fix $\mu$ to the value $\mu_{\rm SSS}$ it takes
on the SSS solutions.  However, there are two difficulties:
First, although $\mu_{\rm SSS}$ is a well-defined function of the isolated
horizon parameters $r_\Delta,{Q}_\Delta$ and $P_\Delta$,
where static solutions exist, there is no
closed expression for it in terms of these parameters simply
because the full set of solutions (including the
colored black holes) is not known in closed form.
The second and more serious problem is that,
for an arbitrary point in parameter space, there might
not be any SSS solution (recall that the colored black holes
span only a countable number of points in the $P_\Delta$ axis for
a given value of $r_\Delta$). As we shall see in next section,
one can still have a consistent Hamiltonian formulation and a first law
even for those isolated horizons lying on points of the parameter
space where no SSS solutions exist. But, as should be clear from
the discussion, at this point there
is no canonical normalization of $\mu$ on the whole space ${\cal IH}$.

Even when we can not find an explicit functional form for $\mu$, we
can still write the general form that $\kappa_{\l}$ shall have.
Using an identity coming from the isolated horizons boundary
condition we have the following relation \cite{abf2},
\be
\kappa_{\l}\= \frac{1}{\mu}\left(-\frac{2\pi}{a_\Delta}+
\Phi_{11}\right)\, .
\ee
Using the expression for $\Phi_{11}$ in terms of the charges (\ref{ricci})
we have,
\be
\kappa_{\l}=-\frac{1}{\mu}\;\frac{1}{2r^2_\Delta}\left[1-
\frac{(Q_\Delta^2+P_\Delta^2)}{r_\Delta^2}\right]\, .
\ee
For those points of parameter space $\Lab$ where an SSS static
solution exists, we can go further and make use of
the general form of the metric (\ref{SSS}). One
then finds that the
expansion of the properly normalized $n$ is such that
\be
\mu_{\rm SSS}=-\frac{e^{\delta(r_\Delta)-
\delta(\infty)}}{r_\Delta}\, .
\ee
A coordinate transformation can always render $\delta(\infty)=0$,
so the properly normalized $\kappa$ for general isolated horizons
takes the form,
\be\label{5.8}
\kappa= \frac{e^{-\delta(r_\Delta,Q_\Delta,P_\Delta)}}{2r_\Delta}
\left[1-\frac{(Q_\Delta^2+P_\Delta^2)}{r_\Delta^2}\right]\, .
\ee
It is in a sense remarkable that our ignorance about the explicit
form of the solutions is encoded in the function $\delta$.

We end this section with a discussion. As we have emphasized in
this section, any choice of $\mu$ in terms of the horizon parameters
defines a vector field $\l^a$ and a surface gravity on ${\cal IH}$.
However, one would like to make contact with the space of static
solutions in such a way that the surface gravity $\kappa_\l$
coincides with the surface gravity of the properly normalized
Killing field. In the case of Einstein-Maxwell-Dilaton \cite{acdil}
this was indeed possible since the static solutions span the space
of horizon parameters $\Lab$. That is, for each value of the
isolated horizon parameters, there exists a (unique) static
solution which allows us to fix $\mu$ in a unique, canonical way.
In the case of EYM, the space of SSS solutions does not span $\Lab$.
Thus, one is able to canonically fix $\kappa$ only in this
subspace; for a general point, there is no preferred normalization.
This seems to be a serious shortcoming of the formalism that
might place the EYM system on a different status than the
EM and EMD systems.
As we shall see in next section, this ambiguity is also
manifested in the definition of mass. However, this problem will
also motivate a second conjecture that we shall put forward
in Section~\ref{sec7}.

\section{Mass and the First Law}
\label{sec6}

The space-times that admit an isolated horizon
are not necessarily stationary, therefore, it is no longer
meaningful to identify the ADM mass $M$ with the mass $M_\Delta$ of
the isolated horizon.  For the formulation of the first law,
we must first introduce an appropriate definition of
$M_\Delta$. This definition should be general enough in the
sense that, for each choice for the normalization of $\l^a$, the Mass
should be uniquely defined, and the first law should be
valid for \textit{any} choice of normalization.
As we shall see, the Hamiltonian framework provides a natural strategy.
In the Einstein-Maxwell-Dilaton case the total Hamiltonian
consists of a bulk term and \textit{two} surface terms,
one at infinity and the other at
the isolated horizon.  As usual, the bulk term is a linear combination
of constraints and the surface term at infinity yields the ADM energy.
In a rest-frame adapted to the horizon it is then natural to identify
the surface term at $\Delta$ as the horizon mass, $M_\Delta$.  Indeed,
there are several considerations that support this identification
\cite{abf2}.

For the gravitational part of the action and Hamiltonian, the
discussion of \cite{abf2} only assumed that the stress-energy tensor
satisfies two conditions at $\Delta$: i) $-T_{ab}\l^a$ is a future
pointing causal vector field on $\Delta$; and, ii) $T_{ab}\l^a n^b$ is
spherically symmetric on $\Delta$.  Both these conditions are met in
the present case.  Therefore, we can take over the results of
\cite{abf2} directly.  For the matter part of the action and
Hamiltonian, the overall situation is again analogous, although there
are the obvious differences in the detailed expressions.
 As in the Einstein-Maxwell case, matter terms
contribute to the surface terms in the Hamiltonian only because one
has to perform one integration by parts to obtain the Gauss constraint
in the bulk term.

The net result is the following.  Consider a foliation of the given
space-time region $\M$ by a 1-parameter family of (partial)
Cauchy surfaces $M_t$, each of which extends from the isolated horizon
$\Delta$ to spatial infinity $i^o$ (see Figure).  We will assume that
$M_t$ intersects $\Delta$ in a 2-sphere belonging to our preferred
foliation and that the initial data induced on $M_t$ are asymptotically
flat. Denote by $S_\Delta$ and $S_\infty$ the 2-sphere boundary of $M_t$
at the horizon and infinity, respectively.  Choose a time-like vector
field $t^a$ in $\M$ which tends to the unit time-translation
orthogonal to the foliation at spatial infinity and to the vector
field $\l^a$ on $\Delta$, with the normalization fixed (or partially
fixed) as in Section \ref{sec5}.  Then,
the Hamiltonian $H_t$ generating evolution along $t^a$ is given by:
\ba \label{ham}
H_t &=& \int_{M_t} {\rm constraints}\,\,+\,\, \lim_{r_o\to \infty}
\oint_{S_{r_o}}\left(\frac{r_o}{4\pi G}\, {\Psi_2}\right) \,\,
{}^2\epsilon -\Phi_{\infty}Q_{\infty} \nonumber\\
&+& \oint_{S_\Delta} \left(\frac{\mu^{-1}}{4\pi G} \Psi_2 \right)
{}^2\epsilon + |(\bA\cdot l)|{Q}_\Delta + V\, ,
\ea
where $S_{r_o}$ are large 2-spheres of radius $r_o$ and
$V$ is a constant on ${\cal IH}_0$, the
space of isolated horizons with fixed values of the
horizon parameters. (The calculation
and the final result are completely analogous to those in the
Einstein-Maxwell case \cite{abf2}.)  Note
that the surface terms depend only on the `Coulombic' parts of the
gravitational and Yang-Mills fields.

It is easy to check that the surface term at infinity is, as usual,
the time component $P_a^{\rm ADM}t^a$ of the ADM 4-momentum $P_a^{\rm
ADM}$, which in the present (-,+,+,+) signature is negative of the ADM
energy, $P_a^{\rm ADM} t^a = -E^{\rm ADM}$.  It is natural to identify
the surface term at $S_\Delta$ as the energy of the isolated horizon.
(There is no minus sign because $S_\Delta$ is the \textit{inner}
boundary of $M$).  Since $t^a \= \l^a$ and since $\l^a$ represents the
`rest frame' of the isolated horizon, this energy can in turn be
identified with the horizon mass $M_\Delta$.  Thus, we have:
\be \label{mass}
 M^{(\l)}_\Delta
= \oint_{S_\Delta} \left(\frac{\mu^{-1}}{4\pi G} \Psi_2 \right)
{}^2\epsilon + |(\bA\cdot l)|{Q}_\Delta +
V_{(\l)}(a_\Delta,Q_\Delta,P_\Delta) \, .
\ee
Here, the so far undetermined function $V$, depends only on the
horizon parameters (and coupling constants). In the variational principle,
this term played no role, but in the Hamiltonian description it
becomes essential, since we are now interested in variations along
the full isolated horizons phase space ${\cal IH}$. Thus, one should
be able to consider in the formalism displacements along directions
in which the horizon parameters change. As we shall show below,
requiring a consistent Hamiltonian formulation enables us
to determine the function $V$ for the EYM system.
Now, using the expression (\ref{4.13}) of surface gravity in terms of the
Weyl tensor
(and $\mu$), and calling $\Phi_\Delta:=|\bA\cdot \l|$ on
$\Delta$, we can cast $M_\Delta$ in a more familiar form:
\be \label{smarr}
M_\Delta = \frac{1}{4\pi}\, \kappa a_\Delta\, +\, \Phi_\Delta Q_\Delta
+ V(a_\Delta,Q_\Delta,P_\Delta)\, ,
\ee
where we have dropped the explicit $\l^a$ dependence of the Mass for
notational simplicity.
Thus, as in the Einstein-Maxwell case, we obtain a Smarr formula.
 However,
the meaning of various symbols in the equation is somewhat different.
Since an isolated horizon need not be a Killing horizon, in general
$M_\Delta$ does \text{not} equal the ADM mass, nor is $\kappa$ or
$\Phi_\Delta$ computed using a Killing field. Since
the constraints are satisfied in any solution, the bulk term in
(\ref{ham}) vanishes as well.  Hence, in this case, $H_t=M_\Delta - E^{\rm
ADM}$, the difference being the `radiative energy'
in the space-time.
Finally, as emphasized in \cite{abf2}, the matter
contribution to the mass formula (\ref{mass}) is subtle: while it does
not include the energy in radiation outside the horizon, it does
include the energy in the `Coulombic part' of the field associated
with the black hole hair. (Recall that the future limit of the Bondi
energy has this property.) This fact is crucial to the analysis of the
`physical process version' of the first law. However, since this issue
was discussed in detail in \cite{abf2}, we shall not discuss this here.

Now, a consistent Hamiltonian formulation
(for a sector of a diffeomorphism invariant theory) requires
 that  for an arbitrary vector $\delta$
tangent to the symplectic manifold, (i.e. the phase space $\Gamma$), one has
\be
\delta H = \Omega (\delta, X_H)\, ,
\label{1}
\ee
where $X_H$ is the vector field that corresponds
to the equations of motion for a
given choice of lapse and shift, and $H$ is the Hamiltonian function
corresponding to the same choice of lapse and shift.
In the preceding  prescription one assumes that the evaluation of
$\delta H$  is carried out by considering the change in $H$ associated
with the displacement $\delta$ of the phase space point, but keeping the
lapse and shift \textit{fixed}. If we now let the choice of
lapse and shift depend on
the phase space point --as is the case when
$\l$ is normalized as in Sec.~\ref{sec5}--
we would obtain a \textit{new} variation
$\delta\tilde{H}$.
This might fail to satisfy
\be
\delta\tilde{H} = \Omega (\delta, X_H)\, ,
\ee
with the \textit{same} $X_H$ as in (\ref{1}).
It turns out that  the necessary and
sufficient condition  to obtain the required consistency is
the validity of the first law\cite{abhay}.
\be
\delta M_\Delta = {1\over 8 \pi }
\kappa \delta a_\Delta + \Phi_\Delta \delta Q_\Delta\, .
\label{firstlaw}
\ee
That is, the first law of black hole mechanics
-for quantities defined only at the horizon- arises
naturally as part of the requirements for
a consistent Hamiltonian formulation
in which, for every value of the horizon parameters,  one
has chosen a canonical
lapse and shift functions making the
latter dependent on the point in phase space. Note  that  in contrast
with the above situation, when constructing a
Hamiltonian to deal with the analogous problem at infinity, the canonical
choice of normalization of lapse and shift at infinity is taken as
independent of the
phase space point, namely, they are chosen to correspond to a unit time
translation (normal to the initial-data hyper-surface) at infinity.

In the cases of Einstein Vacuum, Einstein-Maxwell and
Einstein-Maxwell-Dilaton theories, this consistency requirement
 translate into an
identity that is automatically satisfied by the expressions of $M_\Delta$,
$\kappa$ etc, for all values of the parameters and variations there-off.
In the case of Einstein-Yang-Mills Theory, -as well as in other theories
where hair is present-, the only
way to ensure the validity of the consistency requirement is to limit the
class of variations $\delta$ allowed.
This has two dramatic consequences: First, it
defines a foliation of phase space
by a collection of (symplectic) leaves
over which the Hamiltonian formulation is consistent,
a situation which puts the construction in the `non-standard' class.
In fact, some simple systems are described by a
similar type of situation, as
for example  the (reduced) Hamiltonian description of a rotating body
where the -3 dimensional- phase space
is foliated by two spheres, each of which is a
true symplectic manifold where the Hamiltonian motion is restricted
\cite{arnold,marsden}. Second, the status of the first law changes
from that of an identity, valid for all variations, to that
of a specification of the class of variations that the formalism allows.

It is important to note that in the first law (\ref{firstlaw})
only variations of the electric charge are involved, and not
variations of the magnetic charge. On the other hand, the Horizon
Mass (\ref{smarr}) might depend on $P_\Delta$ through $V$.

Let us now see that asking consistency of the formalism
leads us to
some conditions that the function $V$ should satisfy.
In order to do this we shall follow, for completeness,
Ref. \cite{cb:ac} closely.
The first step in this direction is to regard (\ref{firstlaw})
as an identity between one forms $\d M_\Delta=\frac{1}{8\pi}\kappa
\d a_\Delta + \Phi_\Delta\d Q_\Delta$, where $\kappa, \Phi_\Delta$
and $V$ are
functions on $\Lab$.
Thus one can consider differential forms on $\Lab$ and
take an exterior derivative of the
`first law' to arrive at,
\be
0=\frac{1}{8\pi}\,\d\kappa\wedge\d a_\Delta+\d \Phi_\Delta
\wedge
\d Q_\Delta\, .
\label{cond1}
\ee
The first conclusion
coming from (\ref{cond1}) is that the  variations on
$\Lab$ are restricted to sub-manifolds such
that the pull-back of the form
$\d P_\Delta\wedge\d a_\Delta$ vanishes. That is,
 $P_\Delta$ is not free to vary
independently  of $r_\Delta$ and $Q_\Delta$.
This is precisely what happens
for SSS solutions (representing only a {\it discrete} set of curves
in the plane $(r_\Delta,P_\Delta)$) and in the static axial-symmetric
case \cite{kk} (also covering a discrete set of curves in the
plane).
{}From now on, we restrict ourselves to the symplectic leaves where
the formalism is well defined. On these
sub-manifolds the magnetic charge
becomes a function of the area and electric charge,
$P_\Delta=P_\Delta(r_\Delta,Q_\Delta)$.

Second, the condition (\ref{cond1}) gives us a relation between
$\kappa$ and $\Phi_\Delta$,
\be
\frac{\partial\kappa}{\partial Q_\Delta}
=8\pi\frac{\partial\Phi_\Delta
}{\partial a_\Delta}\, .
\label{cond1.5}
\ee
Thus, if we know the surface gravity $\kappa$ then we can
derive an expression for the potential $\Phi_\Delta$.

Finally, taking the variation of (\ref{smarr}) and comparing it to
(\ref{firstlaw}) we arrive at the following equations,
\ba\label{cond2}
a_\Delta\,\frac{\partial \beta}{\partial a_\Delta}+
8\pi r_\Delta\, Q_\Delta
\frac{\partial \Phi_\Delta
}{\partial a_\Delta}
+8\pi r_\Delta\,
\frac{\partial V}{\partial a_\Delta}&=&0\, ,\\
\frac{r_\Delta}{2}\,\frac{\partial\beta}{\partial Q_\Delta}+
Q_\Delta\,\frac{\partial\Phi_\Delta
}{\partial Q_\Delta}
+\frac{\partial V}{\partial Q_\Delta}&=&0\, ,
\label{cond3}
\ea
where $a_\Delta=4\pi r^2_\Delta$ and, for convenience,
we have defined $\beta:=2r_\Delta\,\kappa$.
Thus, given $\kappa$ and $\Phi_\Delta
$ one can in principle, integrate equations
(\ref{cond2}) and (\ref{cond3}) to find $V$.
Note that these equations are defined over the
horizon parameters space $\Lab$,
so they are completely {\it local} to the horizon $\Delta$.

Recall that the general prescription for arriving at an
explicit  expression for surface gravity $\kappa$, for general
isolated horizons, involves the fixing of the expansion $\mu$
as function of the horizon parameters. For this, one requires some
input from the SSS solutions
(where they exist). However,
it is important to stress that the results of this section regarding
a consistent Hamiltonian formulation and the validity of the first law,
are independent of the particular choice of normalization $\mu$
(and $\kappa$) that one makes.
Thus, there is a consistent Hamiltonian for each choice.
This is particularly important for those points of horizon parameter
space where no SSS
solutions exist and therefore, no `canonical' normalization is available.
Nevertheless, Isolated Horizons still exist,
and are well defined for those points of parameter space.
It is when we want to have a canonical choice of $\mu$ and therefore,
of $\kappa$ and $M_\Delta$, that we are forced
make contact with static solutions
(for the allowed regions in $\Lab$).

In the remaining of this section, we focus our attention
to static spherically symmetric (SSS) solutions
to the EYM equations described in Section \ref{sec2}.
We shall consider the three classes of SSS solutions
described in Section~\ref{sec2},
and find expressions for their
surface gravity $\kappa$ and
Horizon Masses $M_\Delta$. As discussed
before, a study of the three sectors of SSS solutions serves two
purposes. First, it provides us with a
way of fixing the normalization of $\l^a$ for
\textit{general} isolated horizons for those point of parameter
space $\Lab$
where SSS solutions exist, and second, it will allow us
to find, in the next Section, new results regarding SSS solutions.
This is because, even
when the expressions we will find for $\kappa$ and $M_\Delta$
are valid in the general framework, they are, in particular, also
valid for SSS solutions.
Some of the results of this and the next section,
regarding SSS colored black holes, were already reported in
\cite{ac:ds}.

Let us start with the electrically charged case,
corresponding to the $P_\Delta=0$ surface in $\Lab$.
Since these
solutions are nothing but electrically charged Reissner Nordstrom
solutions, the expansion of $n$ is given by \cite{abf2},
\be
\mu=-\frac{1}{r_\Delta}\label{mu}\, ,
\ee
and the surface gravity is given by
\be
\kappa=\frac{1}{2r_\Delta}\left( 1-\frac{Q^2_\Delta}{r^2_\Delta}\right).
\ee
We can now consider Equation (\ref{cond1.5}) and find that
$\partial_{r_\Delta}\Phi_\Delta
=-\frac{Q_\Delta}{r_\Delta^2}$. Then, asking
$\Phi_\Delta
$ to vanish as $r_\Delta \mapsto \infty$, we have that
\be
\Phi_\Delta
=\frac{Q_\Delta}{r_\Delta}\, ,
\ee
which corresponds precisely to the value of the electric potential
on RN solutions.
The equations (\ref{cond2}) now imply that $\partial_{r_\Delta} V=0$
and $\partial_{Q_\Delta} V=0$. Since the restriction of
 $V$ to the plane $P_\Delta=0$ does not depend on $P_\Delta$,
the only possibility is that $V={\rm constant}$ on the
$P_\Delta=0$ plane of $\Lab$. In
order to fix the value of $V$ we notice that the Schwarzschild
one-parameter family of solutions
--corresponding to zero electric field-- are contained
within the electric RN family. These solutions
are purely gravitational since the gauge potential vanishes exactly, and
it is known that for the pure Einstein theory one can set $V=0$
(see \cite{abf2}).

We now have expressions of the mass $M_\Delta$, surface gravity
$\kappa$, area $a_\Delta$ and the electric potential $\Phi_\Delta
$ of any
isolated horizon in terms of its fundamental parameters $r_\Delta,
Q_\Delta,$:
\be
M_\Delta =\frac{r_\Delta}{2}\left[
1+\frac{Q^2_\Delta}{r_\Delta^2}
\right]\, .
\label{7.4}
\ee
This is precisely the same form as in the Einstein-Maxwell theory.
The total energy of the system $E$, related to
on-shell value of the Hamiltonian,
is given by
\be
E=-H_t=M_{\rm ADM} - M_\Delta\, ,
\label{7.5}
\ee
which for the electric RN embedded family vanishes exactly.

Let us now consider embedded Abelian magnetic solutions. For this
solutions, the electric charge $Q_\Delta$ vanishes so the Horizon mass
variation formula, when restricted to the purely magnetic
sector of the SSS space takes the form,
\be
\delta M_{\Delta}=\frac{1}{8\pi}\,\kappa\,\delta a_\Delta\, .
\label{22b}
\ee
The
formula for the mass (\ref{smarr}) is given by,
\be
M_\Delta=\frac{1}{4\pi}\kappa \;a_\Delta +
V(r_\Delta,Q_\Delta=0,P_\Delta=1),
\label{12b}
\ee
The normalization factor $\mu$ remains the same as in the electrically
charged case given by (\ref{mu}) and the surface gravity is given by
\be
\kappa=\frac{1}{2r_\Delta}\left( 1-\frac{P^2_\Delta}{r^2_\Delta}\right).
\label{kappamag}
\ee
Then equation (\ref{cond1.5}) implies that $\Phi_\Delta
$ is zero. The second set
of equations (\ref{cond2}) and (\ref{cond3}) reduce to,
\be
\partial_{Q_\Delta}V = 0\, ,
\label{7.9}
\ee
and
\be
\partial_{r_\Delta} V=-\frac{r_\Delta}{2}\partial_{r_\Delta}\beta\, .
\label{7.10}
\ee
Using equation (\ref{kappamag}) we get the following equation that
$V$ should satisfy,
\be
\partial_{r_\Delta}V=-\frac{P^2_\Delta}{r^2_\Delta}\, .
\label{7.11}
\ee
Now, recall that, for a given value of the magnetic charge
$P_\Delta$, one has black hole solutions for
$r_\Delta\geq |P_\Delta|$ (the extreme case corresponding
to $r_\Delta = |P_\Delta|$). In order to integrate (\ref{7.11}),
one has to choose some `boundary conditions' on the space $\Lab$.
Our choice, motivated by consistency with the colored black holes
(see below), is to set $V(r_\Delta=P_\Delta)=0$. With this choice,
$V$ takes the form,
\be
V=\int^{r_\Delta}_{|P_\Delta|}\frac{P^2_\Delta}{\tilde{r}^2}\d\tilde{r}=
\frac{P_\Delta^2}{r_\Delta}- |P_\Delta|\, .
\ee
With this, the Horizon Mass $M_\Delta$ is given by,
\ba
M_\Delta&=&\frac{r_\Delta}{2}\left(1+\frac{P^2_\Delta}{r^2_\Delta}
\right)- |P_\Delta|\, ,\nonumber\\
&=& M_{\rm ADM} -|P_\Delta|\, .\label{7.13}
\ea

Let us now consider the case in which the Abelian solution has both
electric and (unit) magnetic charge. The surface gravity is given by,
\be
\kappa=\frac{1}{2r_\Delta}\left[
1-\frac{(Q^2_\Delta+P^2_\Delta)}{r^2_\Delta}\right]\, .
\label{6.24}
\ee
Equation (\ref{cond1.5}) leads us to conclude that
$\Phi_\Delta=\frac{Q_\Delta}{r_\Delta}$, and equation (\ref{cond2}) takes
the form,
\be
\partial_{r_\Delta}V=-\frac{(Q_\Delta^2+P_\Delta^2)}{r_\Delta^2}+
\frac{Q_\Delta^2}{r_\Delta^2}=- \frac{P_\Delta^2}{r_\Delta^2}\, .
\ee
Then, imposing again the boundary condition that $V$ vanishes on
extremal solutions. we have that,
\be
V=\frac{P_\Delta^2}{r_\Delta^2}-\frac{P_\Delta}{\sqrt{P^2_\Delta+
Q^2_\Delta}}\, .
\ee
The Horizon Mass is now,
\be
M_\Delta=M_{\rm ADM}-\frac{P_\Delta}{\sqrt{P^2_\Delta+
Q^2_\Delta}}\, ,
\label{6.27}
\ee
and the total energy is then $E=\frac{P_\Delta}{\sqrt{P^2_\Delta+
Q^2_\Delta}}=\frac{1}{\sqrt{1+Q_\Delta^2}}$.

Finally, there is the
most interesting case, i.e., the family of colored black
holes labeled by $r_\Delta$ and an integer $n$.
Since these solutions correspond to the purely magnetic case,
Equations (\ref{22b}), (\ref{12b}), (\ref{7.9}) and
(\ref{7.10}) continue to hold.
This last condition that the function $V=V(r_\Delta)$
should satisfy can be written as,
\be
V^\prime=-\frac{r_\Delta}{2}\beta^\prime\, ,
\ee
with `prime' denoting differentiation with respect to $r_\Delta$.
(We are considering variation with fixed value of $n$.)
Furthermore, by requiring that $M_\Delta\mapsto 0$ as
$r_\Delta\mapsto 0$, -coming from
physical considerations-
we arrive at the following relation,
\be
M_\Delta=\frac{1}{2}\int_0^{r_\Delta} \,\beta(\tilde{r})\,
\d \tilde{r}\, , \label{13}
\ee
where the integration is again performed over the {\it space of parameters}
of the $n$-colored black hole,
labeled by the horizon radius $r_\Delta$, and
not over space-time.
Let us note that for the $n=0$ solution, where $\beta$ is known in
closed form ($\beta=1$), we arrive at $M_\Delta^{(n=0)}=r_\Delta/2=
\kappa a_\Delta/(4\pi)$, as expected.

Several remarks are in order. First, we must emphasize that the
determination of $V$, and thus of $M_\Delta$ relied on considerations
involving only variations of quantities associated with the horizon
$\Delta$. Thus, the horizon mass is a well defined quantity
in the isolated horizons phase space ${\cal IH}$
(provided that a global normalization of $\mu$ exists).
Second, the HHM
defined by (\ref{smarr}), when restricted to SSS configurations,
does not agree with the usual definitions of mass that one finds in
the literature (see for instance \cite{review} and \cite{heusler2}).
It should be stressed that (\ref{13}) comes from a consistent
Hamiltonian formulation, and is {\it not} a definition as occurs
in other treatments.

As it was discussed at the end of Section~\ref{sec5}, the fact that
SSS solutions to the EYM equations do not span the space
$\Lab$ of horizon parameters is, in a sense, disturbing. It might seem
that the Einstein-Yang-Mills system is in a different status than
the Einstein-Maxwell-dilaton (EMD) system where the space of
horizon parameters is in a one to one correspondence with the
space of static solutions.
One would like to have a similar result in the EYM case. However,
the SSS solutions span only a subspace of $\Lab$. Luckily, the
Spherically Symmetric solutions in EYM do not exhaust all possible
Static solutions (as occurs in EMD); there are static solutions
with axial symmetry that are not spherically symmetric \cite{kk}.
With these results at hand, we propose a completeness conjecture
in the next Section.

\section{The Canonical Normalization Problem: A Proposal}
\label{sec7}

We have seen that in order for the
isolated horizons scheme to define the
surface gravity and the horizon mass
of the colored black holes we needed to introduce a uniqueness
conjecture $C1$
that guarantees that, given the isolated horizon parameters
$a_\Delta, P_\Delta, Q_\Delta$,
there would be at most
one SSS solution. This was needed for, otherwise, the normalization
of $\mu$ would not be
uniquely specified  given those parameters. The existing numerical
evidence does indeed
strongly support this conjecture. However,
as we have mentioned before, and
as is evident from the previous discussion, this is not sufficient
in  order to have
the Isolated Horizon framework  working for the EYM system to the same
extent that it works,
say, for the Einstein Vacuum, Einstein-Maxwell, and
Einstein-Maxwell-Dilaton systems.
In order to achieve that, we would need to have a canonical
normalization of $\mu$ for a `complete'
set of values of the Isolated Horizon parameters. In the
previously mentioned cases this canonical choice is given by
the existence of
static (and spherically symmetric) Black Hole  solutions for all
isolated horizon  values of the parameters.%
\footnote{
In fact, although this is true for the Einstein Vacuum system,
in the Einstein Maxwell case we already
have a potential problem, because if $Q>r_\Delta$ there are
no such static solutions. The consistency of the
whole scheme would require the impossibility of constructing a
space-time containing an Isolated Horizon
with such values of the parameters. This is a rather serious
consequence of the present point of view, and one that should be testable.
There is a strong correlation of these issues with the
cosmic censorship conjecture, that prevents us from violating
the inequality $Q<r_\Delta$ for static solutions. It
 would seem that the IH formalism  implies that one can
not construct initial data for a solution containing
a black hole with  values of the
parameters that violate this inequality.}

There is strong numerical evidence against the validity of the analogous
claim
in the case of the EYM system. In fact in the regime of staticity
and spherical symmetry there are,
given a fixed value of $a_\Delta$, only a
discrete set of values of $P_\Delta$ for which there are black
hole solutions.
Moreover, within this regime there are
 no  Black Hole solutions for any value of
$P_\Delta \not = 1, 0$ and $Q_\Delta \not =0$. Thus if we want to have
any hope
that any claim in that direction might be true,
 we must formulate it outside this
restrictive regime.
Indeed the fact that in EYM systems there are static Black Hole solutions
that are not spherically symmetric, (indicating that
that the analogous to Israel's theorem is false in this case),
already shows us that
we must go beyond the SSS regime. In fact the solutions alluded
above are axially symmetric, instead of spherically symmetric,
but seem to
share, with the SSS solutions, the discreetness of the allowed
values of
$P_\Delta$ (at least to the extent that this issue has been studied
\cite{kk}).
Thus we have to go beyond this regime as well.  In fact
there are strong indications (see for example
  the discussion in \cite{sud:wal}) that we must go
beyond the static regime, and pose the conjecture in a broad enough setting
that would still allow one to single out, for a
given choice of IH parameters, a particular black hole
solution and thus a canonical normalization of $\mu$.
This would be of course the class of stationary black hole solutions,
where we would have to keep track also of the angular momentum,
both at infinity $J_\infty $ and at the horizon $J_\Delta$. The
completeness conjecture would thus be:
$C2$:
{\it For every value of the Isolated Horizon parameters
$a_\Delta, P_\Delta, Q_\Delta, J_\Delta$ for which a space-time
can be constructed, there exist also a  stationary Black Hole Solution
with the same value of the parameters, now characterizing
the Killing Horizon}.%
\footnote{In this statement, a space-time
`can be constructed' whenever there exists an asymptotically
flat solution to the EYM equation satisfying IH boundary conditions
with the specified values of the horizon parameters.}

Let us now consider some of the implications of this conjecture.
First, a stationary
black hole solution would be characterized by its parameters at infinity:
 $M_{\rm ADM}, P_\infty, Q_\infty, J_\infty $ and therefore
 the conjecture would
imply the existence of a well defined map
$ \Psi  :(a_\Delta, P_\Delta, Q_\Delta, J_\Delta ) \rightarrow
(M_{\rm ADM}, P_\infty, Q_\infty, J_\infty)$ The failure of the no
hair conjecture
would indicate that this map is not invertible.
In fact we know that it would  not be injective. Moreover the map
would be nontrivally four dimensional,
in the sense that fixing, say, $J_\Delta =0$
would not fix $J_\infty=0 $, as can be seen from the
following expression,
\cite{sw2}
\be
4\pi\Omega(J_\Delta-{J}_\infty)=\int_{\Sigma}\d^3x \,\left(
t^b\tilde{\bE}^a_i
\bF^i_{ab}+{\cal L}_t(\bA^i_a)\tilde{E}^a_i\right)\, ,
\label{sw:a}
\ee
valid for stationary black hole solutions in EYM theory.
Here, $\Sigma$ is a maximal hyper-surface intersecting
the bifurcate horizon, and $t^a$ is the projection 
to $\Sigma$ of the time translation Killing field. 
As it can be seen from the Eq. (\ref{sw:a}),
there is a bulk  contribution to ${J}_\infty$, the canonical
angular momentum at infinity.
Here, $J_\Delta$ is a particular definition
of `horizon angular momentum'
(given by a Komar integral), and
$\Omega$ stands for the angular velocity
of the horizon, i.e, the expression appearing in the first law
\be
\delta M_{\rm ADM}+\Phi_\infty\delta Q_\infty-\Omega\delta
J_\infty=\frac{1}{8\pi}\kappa\delta a_\Delta\, ,
\label{sw:b}
\ee
where $\Phi_\infty$ is the `electric potential' at infinity.
In fact the EYM system is, in this respect, rather different from the
Einstein-Maxwell system, because in the latter one can disentangle for
example the expression for $\Phi_\infty Q$
from the expression for $\Omega J_\infty$, something that can not be
done in the former\cite{sw2},
in which case the only relationship that can be obtained is
given by the expression,
\be
8\pi(\Omega J_\infty-\Phi_\infty Q_\infty)=
\int_\Sigma N(\pi_{ab}\pi^{ab}+2\bE^a_i\bE_a^i)/h\, ,
\ee
with $\pi^{ab}$ the momentum conjugate to $h_{ab}$, the Riemannian metric
on $M$ and $N$ the `lapse function'.

Assuming the validity of $C2$ one would have a canonical
choice for the normalizations that could be used to uniquely define
$\kappa$ and $M_\Delta$,
the one provided by the Killing field of the stationary
solution that is null at
the horizon and that is normalized so that at infinity is a
unit time translation.
In order to make all these considerations more precise from the isolated
horizons point of view, one needs to consider the extension of the
formalism given in \cite{afk} and \cite{abl}.

 Now,
let us concentrate for the moment in the SSS sector and see if
we can understand
the discreteness observed there in terms of this conjecture.
In the analysis that lead to the discovery of  SSS  Black Holes
in EYM theory, one is fixing
$P_\infty=0$ because one is interested in solutions that are
Abelian at large distances. Moreover spherical symmetry
evidently requires
$J_\infty =0$ and $ J_\Delta =0$.  Moreover the mixture alluded
to before would
prevent us from achieving this (spherical symmetry)
 unless we also set $Q_\infty =0$.
Thus we see that the (highly nonlinear) problem is given
by four constraints in a four dimensional space, and thus that the
set of solutions is expected to be given by a discrete set
(i.e. the linearized problem about a given solution in the
SSS sector has no
nontrivial solution within the sector).

Next, let us
consider some consequences of the completeness conjecture
in the structure of the space $\Lab$. We have seen in Sec.~\ref{sec6}
that the space ${\cal IH}$ (where we now
have to include distortion and rotation \cite{afk,abl})
is foliated by `symplectic leaves' where
the Hamiltonian formulation and the first law are valid. This foliation
intersects the space ${\cal S}$ of stationary solutions and defines
a one parameter foliation of it. Now, if the $C2$ conjecture is valid,
we have an isomorphism between ${\cal S}$ and $\Lab$, which
then induces a canonical foliation of $\Lab$.

On the other hand,  we must point out the following heuristic argument
against the conjecture. Consider a  stationary black hole
solution in EYM theory, in order to be asymptotically flat the YM field
strength must fall off rather rapidly at large distances, thus
the self interaction of the fields must be falling off faster
than the fields themselves and thus the fields must behave in the large
distance limit as free fields, i.e., as Abelian fields. This suggests that
the only possible values of the magnetic charge at $\infty$,
$P_\infty$ are the
Abelian values $0,1,..$ (there is of course no such restriction on the
allowed value of the electric charge). This view is supported by the
experience
with the static spherically symmetric solutions. Thus, any stationary
solution would be characterized at infinity by the parameters
$M_{\rm ADM}, Q_\infty, J_\infty $ (setting for the moment $P_\infty=0$
to simplify the discussion). On the other hand the solution will be
characterized by its horizon parameters amongst them $a_\Delta$. We know
from the first law in its asymptotic infinity version Eq.
(\ref{sw:b}) (see the discussion in \cite{sud:wal}), that the
stationary black hole solutions are extrema of $M_{\rm ADM}$ at fixed
$a_\Delta, Q_\infty, J_\infty $ within
the constrained phase space (i.e., the space of allowed initial data, which
as usual, can be identified with the space of solutions).
Each such extrema is an isolated point in
that space, so the manifold of stationary solutions is three dimensional.
In \cite{sud:wal} an argument is given that indicates that
there would be a countable infinity
of solutions for each value of the parameters $a_\Delta,
Q_\infty, J_\infty $ which would generalize what happens in
the case of the static solutions. This suggest that the manifold of
stationary solutions is made of a countable infinity of connected three
dimensional components. On the other hand, the conjecture would seem to
indicate that such manifold must be four dimensional. The only way to avoid
this
would require the impossibility of constructing solutions of the equations
representing asymptotically flat black hole space-times (not necessarily
stationary)
containing isolated horizons  for all values of the isolated horizon
parameters.

One would be tempted to take such position in view of the discretness of
the SSS colored black holes with a fixed value of he horizon area
$a^0_\Delta$. Let  $\lbrace{P^i_\Delta}\rbrace_{i=1}^\infty$ be the values
of the horizon magnetic charges of the SSS colored black holes with
horizon area $a^0_\Delta$.
One might want to argue that given a value of
$P'_\Delta \in (0,1) $  that is not in that list, and is,
 say, between two of the values
in the list, one can not construct a solution
representing asymptotically flat black hole space-time  with
 isolated horizon and, say
$a_\Delta=a^0_\Delta, P_\Delta=P'_\Delta, Q_\Delta=0, J_\Delta=0$.
Unfortunately this claim is evidently false: The recipe for constructing
such a space-time is to  give initial data that, upon evolution would be
static
near the horizon and near infinity at least for a finite ``time" interval.
Take the equations for a
the SSS (Eqs. (10),(11) and (12) in \cite{bizon}) 
and set $r_H = \sqrt{a^0_\Delta/4 \pi}$,
 $w(r_H) =\sqrt{1-P'_\Delta}$, and evolve the elliptic equations up to say
$r = 2r_H $ (if we continue to evolve the equations attempting to obtain a
static solution we would find that $w$ diverges so the solution would not
be asymptotically flat). For, say, $r \ge 5r_H$ take the initial data
for the Schwarzschild
solution. In the intermediate region $r \in (r_H, 5r_H)$ take  any
interpolating function for $w$, and set the time derivatives of the
functions $w, m, etc$ (i.e  the ``momenta") to  satisfy the constraints.
The point is that the evolved space-time  will be static in a neighborhood of
the
horizon so, in particular, the Horizon will be isolated, at least during
some finite time interval (until the radiation coming from the
intermediate region arrives at the horizon). Note that,
generically,  in this case the
event horizon will fail to coincide with the isolated horizon.

The argument above suggest that the manifold of isolated
horizon  parameters $\Lab$ is indeed four dimensional and thus the conjecture
would require the identification of this four dimensional manifold with
the infinite set of three dimensional components that seem to constitute
the manifold of stationary solutions. The best that can be hoped at this
point is that the latter be dense in the former, a situation that would
indicate that in the EYM theory there is much  richer structure that
what is found in, say, the Einstein-Maxwell system. In this case, the
canonical normalization for $\l$ would be given by an appropiate
limit (within ${\cal S}$) of solutions where the normalization exists
(assuming, of course, that there is such limit).

On the other hand the argument above is by not means a  tight proof,
particularly so in the case of
the conclusion about the Abelian nature of the allowed values of
$P_\infty$.  As
we have mentioned before, the validity of this conjecture, or some version
of it  (as for example a version based on the assumption that the manifold
of stationary solutions is mapped into a dense 
subset of the isolated horizon parameters),  seems
to be the only
reasonable way in which the Isolated Horizon scheme can be as successful
in the general case as it has proven to be in the Einstein Vacuum,
Einstein-Maxwell and  Einstein-Maxwell-Dilaton theories.

Needless is to say that further research is required in order to elucidate
whether one of the scenarios considered above is correct or whether, in fact,
the Isolated Horizon scheme fails to achieve in EYM theory the same degree
of success that is attained in the previously treated cases.

\section{Spherically Symmetric Static Solutions: Mass and Hair}
\label{sec8}

In this section we shall restrict our attention to the SSS sector of
isolated horizons.
One issue that has been considered for non-Abelian gauge theories is
the relation that might exist between the existence of regular
static, solitonic solutions and  `hairy' black hole solutions.
This issue has been considered, for example, in \cite{bizon2}
from heuristic and dimensional arguments.
In this section, two main issues are studied.
First, by restricting
the Hamiltonian formulation for Isolated black holes
to the SSS sector, we can
define the Hamiltonian Horizon Mass (HHM)
of SSS black holes in EYM theory.
We then use this expression to show that
this quasi-local definition  together with
some basic properties of Hamiltonian Mechanics lead us to a formula
relating HHM and ADM mass of the colored
BH solutions with the ADM mass of the Solitons of
the theory. We also conclude that the
positivity of the `total energy' spectrum of the colored black
holes is related to their instability.

These results are quite surprising, because the IH formalism
was developed to extend the notion of black holes
to situations where radiation is present --and goes
out to infinity-- and one might
have not expected to obtain new results already
in the static sector of the theory.

As we have previously mentioned, the HHM Mass $M_\Delta$ of
a SSS BH does not correspond to any of the Quasi-local definitions
of BH Mass considered in the literature. Furthermore, $M_\Delta$
has the virtue of being constructed from a consistent Hamiltonian
formulation which places it on a different status as the
standard definitions.

To begin, let us calculate the value of the `total energy' $E$ of the
system, for the three sectors of SSS solutions. In order to do this,
we  use a general argument from symplectic geometry
that states that, within each connected component of the space
of static solutions ${\cal S}$
embedded in the space of isolated horizons, the value of the
Hamiltonian $H_t$ remains constant \cite{abf2}.
Let us review this argument
since it is essential for our discussion.
The Hamilton equations of motion
can be written as $\delta H=\Omega(\delta,X_H)$, where $\Omega$
is the symplectic form, $\delta$
is an arbitrary variation and $X_H$ is the Hamiltonian vector field.
A static solution is one at which the Hamiltonian vector field either
vanishes or generates pure gauge evolution. In either case,
the symplectic structure evaluated on $X_H$
and {\it any} arbitrary vector field
$\delta$ vanishes. Therefore, for this point of the phase space,
$\delta H=0$ for any direction $\delta$. In particular $\delta H=0$
for variations relating two static solutions. Now, in the case of
Einstein-Maxwell theory, the no-hair theorems ensure that
all static solutions are given by the RN family. That is,
the space of static solutions is in that case, connected. Furthermore,
since there is
no energy scale in the theory, the only `preferred' value for $H_t$
is zero \cite{abf2}.

What is the situation  in Einstein-Yang-Mills theory? First,
there is the Abelian family of electrically
charged solutions, that represent a connected
component, parameterized by $M,Q$. For these solutions, the basic
reasoning of \cite{abf2} applies and, as follows from (\ref{7.4}) and
(\ref{7.5}),  one has to conclude that
for these solutions $H=0$. However, there is a
 subtle
modification in the case of  magnetic Abelian solutions.
These solutions represent a disconnected component labeled by
one parameter, namely,
the mass $M$ (the magnetic charge $P$ is fixed to be unity).
As discussed in Section~\ref{sec2},
the EYM system possesses an energy scale given by the
YM coupling constant, so in principle,
non-zero values of $H$ are allowed.
In the one-dimensional component corresponding to Abelian magnetic
solutions, the value of $E$ can be
computed using (\ref{7.5}) and (\ref{7.13})
and is given by $E=M_{\rm ADM}-M_{\Delta}=
|P_\Delta|=1$. (We are taking the YM coupling constant $g_{\rm YM}=1$.)

Finally, let us consider colored black holes.
Each connected component of the space of SSS
colored black holes is one-dimensional (parameterized by $a_\Delta$), and
solutions corresponding to distinct values of $n$ belong to
disconnected  components.
That is, the space SSS has a countable number of
connected components. As we shall now show,
for $n\geq 1$ the value of the Hamiltonian turns out
to be {\it different} from zero: $H_t^n\neq 0$.

Recall that the general argument described above tells us that the
(on shell) value of the Hamiltonian is constant for each family
labeled by $n$. This in particular implies that its value is independent of
 the radius $r_\Delta$ of the horizon. Thus one is allowed to take the limit
\be
H^{(n)}=\lim_{r_\Delta\mapsto 0} [M^{(n)}_{\rm ADM}
(r_\Delta)-M^{(n)}_{\Delta}(r_\Delta)]\, .
\ee
Now, it is known that the colored black holes converge point-wise to the
Bartnik-McKinnon soliton solutions \cite{BK} and that the ADM mass satisfies
$M^{(n)}_{\rm ADM}\mapsto M^{(n)}_{\rm BK}$
when $r_\Delta \mapsto 0$. Furthermore,
the horizon mass of the black hole  $M_\Delta$ goes to zero
in this limit, so we can  conclude that
\be
H^{(n)}=M^{(n)}_{\rm BK}\, ,
\ee
that is, the total value of the Hamiltonian equals the mass of the $n$th
Bartnik-McKinnon soliton solution!

\begin{figure}
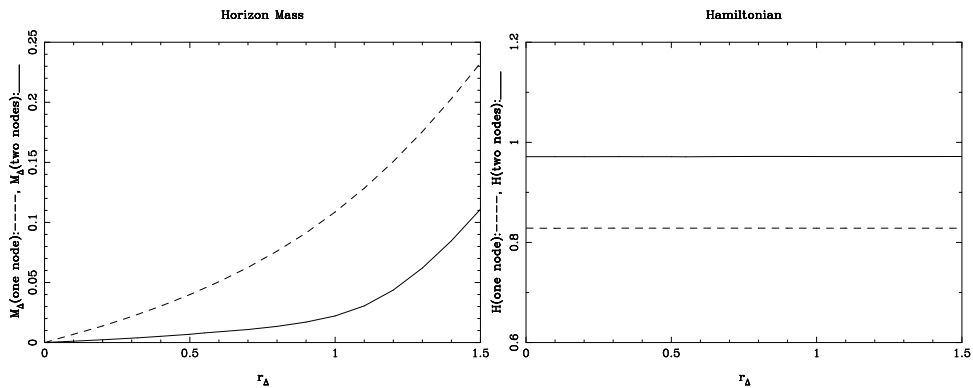

\centerline{\hbox{\psfig{figure=mh.ps,height=5cm,angle=270}}%\hskip=1cm
\hbox{\psfig{figure=H.ps,height=5cm,angle=270}}
}
\bigskip
\caption{The Hamiltonian Horizon Mass $M_\Delta$ and the total Hamiltonian
$H$ are shown as  functions of
the horizon radius $r_\Delta$, for the $n=1,2$ colored black holes
}
\label{hormas}
\end{figure}

\begin{figure}
\centerline{\hbox{\psfig{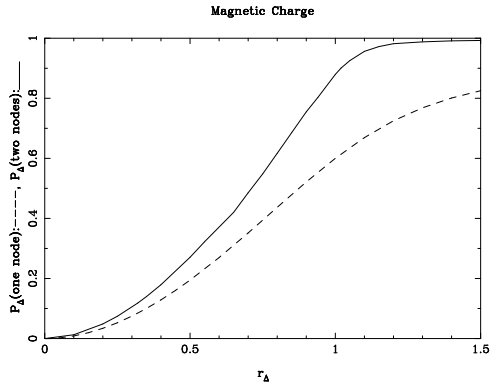}}}
\bigskip
\caption{The Horizon Magnetic Charge $P_\Delta$
is shown as a function of
the horizon radius $r_\Delta$, for the $n=1,2$ colored black holes
}
\label{magnetic}
\end{figure}

We now collect our results for colored black holes
and arrive at the following unexpected relation,
\be
M^{(n)}_{\rm ADM}(r_\Delta)
=M^{(n)}_{\rm BK} +
M^{(n)}_{\rm \Delta}(r_\Delta)\, .
\ee
Thus, we are in the position of writing
an explicit formula for the ADM mass
of the $n$ colored black hole as function of $r_\Delta$,
\be
M^{(n)}_{\rm ADM}(r_\Delta)
=M^{(n)}_{\rm BK} +
\frac{1}{2}\int_0^{r_\Delta} \,\beta^{(n)}(\tilde{r})\,
\d \tilde{r}\, .
\label{adm2}
\ee
In Figure~\ref{hormas}, we show the values of the HHM as a function
of the horizon radius $r_\Delta$, for the $n=1,2$ families. Note that
for a given value of the horizon area, the higher $n$ is, the lower the
horizon mass of the corresponding black hole. In Figure~\ref{magnetic}
the value of the horizon magnetic charge $P_\Delta$ is shown as function
of the radius.

It is important to stress that, a priori,
one would not expect to get the value
of quantities defined at infinity, like the difference of ADM masses
in terms of purely local quantities at $\Delta$.
At this point one might raise the following objection to
the construction of $M_\Delta$ for colored black holes: If we
start by considering the equation
$\delta M_\Delta=\kappa \delta a_\Delta/4\pi$,
and try to integrate it along the one-dimensional curve defined for
each $n$, one can ``trivially" do so by using the
usual form of the first law at infinity, which tells us that the
general solution for $M_\Delta$ is given by $M_\Delta =M_{\rm ADM}
+c$, with $c$ a constant. Then one might argue that the `only' thing
one is doing is to set $c$ such that $M_\Delta(r_\Delta=0)=0$. Thus,
one would conclude, the derivation is trivial and even the notion of
a Horizon mass would seem questionable. This argument would be
perfectly valid had we {\it postulated} the equation
$\delta M_\Delta=\kappa \delta a_\Delta/4\pi$. However, the non-trivial
point here is that this equation comes as a consequence (and consistency
requirement) from a Hamiltonian description respecting --physically
motivated-- boundary conditions. Therefore, even when the algebraic
manipulations are simple, the final result is highly non-trivial.

We can now try to understand the physical meaning of the relation
(\ref{adm2}).
Two facts are known about these solutions:
first, we know that for fixed $a_\Delta$ these solutions represent saddle
points of the ADM mass function $M$ \cite{sud:wal}, and
thus, as one can expect,
for all values of $n$ these solutions are unstable under small
perturbations \cite{uns}.
Let us now note that for the reported solutions in the literature
(see, for instance, \cite{BK}),
the BK Mass is a monotonic function of $n$, starting at
$M_{\rm BK}^1\approx 0.828$ and approaching $1$ as $n$
grows (in standard normalized units).
The fact that the mass of the
soliton, and therefore the total energy of the colored
black holes is positive, confirms our expectation,
coming from energetic considerations, that in general
 $M^{\rm ADM} \geq M_\Delta$.
Indeed, since the difference between the Horizon  and ADM
masses can be seen as the
energy that is available for radiation to fall both into
the black hole and to infinity, one can understand the nonzero value of the
Hamiltonian as an indication that there is a potentiality for instability
of the solution. In static solutions there is, of course, no radiation.
Thus, to be precise, a positive value of the energy $E$ means that,
if we perturb slightly the initial data of a (unstable) static
solution in such a way that the total energy is `very close' to $E$,
then the resulting space-time will approach Schwarzschild in the future,
and the total radiated energy to both infinity and the
horizon will be equal to $E$.
In conclusion, a necessary condition for the solution
to be unstable is for the value of the total energy on the
solution in question to be positive.

Let us conclude this section with four remarks.

\begin{enumerate}

\item %choice of V
In the computation of the Horizon Mass for magnetically charged
Abelian solution, we required that the function $V$ vanishes for
the extremal black holes. Let us now motivate this choice.
first, it is known
that for $r_\Delta\geq 1$, the $n$ colored black holes approach
the Reissner Nordstrom magnetic solution (in a
region around the horizon that expands unboundly
with $n$) when $n\mapsto \infty$.
Let us abuse notation for a moment
and refer to these limiting solution as the `$n=\infty$ colored
black hole'. From the numerical solutions reported in the
literature one can notice that the Horizon Mass
$M^{(n)}_\Delta=M^{(n)}_{\rm ADM}-M^{(n)}_{\rm BK}$ tends to
zero when $n\mapsto \infty$ and $r_\Delta\mapsto 1$ from above.
Now, if we require continuity from
above for $M_\Delta$ on the space $\Lab$ of SSS
solutions, we should require that
\be
\lim_{r_\Delta\mapsto 1^+}
M^{(\infty)}_{\Delta}=0.
\ee
But the Horizon mass is given by $M^{(\infty)}_{\Delta}=
\frac{\kappa a_\Delta}{4\pi}+V$, and $\kappa \mapsto 0$ as
$r_\Delta\mapsto 1$ (since this corresponds to the extremal
RN case). Thus, we conclude that $V\mapsto 0$ when
$r_\Delta\mapsto 1$.

\item %possible tension
Let us now return to the puzzle that we posed in Section~\ref{sec2}:
how can we reconcile the treatments of static solutions
at infinity and
at the horizon? That is, when formulating the
first law, say at infinity, we only have two free parameters
given by the ADM mass and electric charge $Q_\infty$.
However, at the horizon we have three parameters, the
horizon area $a_\Delta$, the electric
charge $Q_\Delta$ and the magnetic charge $P_\Delta$.
Now, equation (\ref{firstlaw})  tell us that their
respective variations depend only on the horizon area and electric
charge, which
seems to be a contradiction. The proper setting of the
problem again comes from the
consistency of the Hamiltonian formulation.
Recall that from Equation (\ref{cond1}) we concluded that the magnetic
charge $P_\Delta$ is not free to vary at the horizon, but its variations
are related to the variations of the horizon radius $r_\Delta$.
This means that the allowed values of $P_\Delta$ and $r_\Delta$
lie in sub-manifolds of co-dimension one embedded in the
space $\Lab$. Thus,
even when we can in  principle have three independent parameters at
the horizons, in practice
when employing the Hamiltonian formulation, one is restricted to
values of the magnetic charge that are determined by the horizon area
(within each family labeled by $n$).
 Thus, the variations of the parameters both
at the horizon and at infinity are consistent.
This can also be seen in the following way.
The results of different  formalisms  at infinity
have shown that
the ADM mass varies as \cite{sud:wal,heusler2},
\be
\delta M_{\rm ADM}=\frac{1}{8\pi}\,\kappa\,\delta a_\Delta +
\Phi_\infty\,\delta Q_\infty\, ,
\label{20}
\ee
but since we know that, at a static solution, an arbitrary variation
$\delta$ satisfies
$\delta E=\delta(M_{\rm ADM}-M_{\Delta})=0$,
we have complete agreement with (\ref{firstlaw}).

Now lets turn our view to the conflict that our completeness conjecture
$C2$ seems to face in view of the
fact that the colored black hole solutions,  with different value of $n$,
have different values of the Hamiltonian and the general argument, presented
in Sec~\ref{sec6},
ensures that $H$ must be constant over any sub-manifold of static
(stationary) solutions.
Thus, this general argument would lead us to conclude that $H$ is
constant over ${\cal S}$.
The solution of the apparent paradox lies in the fact, already mentioned
in Sec~\ref{sec6}
that the consistency of the Hamiltonian formulation leads to a foliation
of phase space by symplectic  leaves over which the
Hamiltonian
formulation is valid. These leaves intersect the manifold
of stationary configurations, which is embedded in ${\cal IH}$.
The intersection results in
hyper-surfaces of constant value of the Hamiltonian.  In the case of
SSS each one of those corresponds to a single  family labeled by
a fixed value of $n$. If we now restrict our consideration to the
hyper-surface with $Q=0$ each of these families  correspond to a curve in
the $(r_\Delta, P_\Delta)$ plane.
Namely, $ P_\Delta$ becomes  a function of
$r_\Delta$ rather than an independent parameter. This again is similar to
what happens in  the case of a rotating body, where the manifold of
stationary states  does not correspond to a single value of the
Hamiltonian,
 but the intersection of a symplectic leave with this manifold
coincides with the curves of constant value of the Hamiltonian.
Then, even when the space $\Lab$ if foliated by a `continuum' of
leaves, the allowed Hamiltonian motions are restricted
to lie within each of these level surfaces of the Hamiltonian
$H$.  This overall picture of the structure of the parameter space is
in fact, consistent with the situation in EMD. In this case, there is only
one `leave' and thus, the intersection with the manifold of static solutions
has also only one leave;
the value of $H$ is indeed constant on the
whole ${\cal S}$ (it is in fact zero).

It is intriguing to note that there seems to be a deep relation between
the existence of non-trivial solitons and
hairy black holes for which the charges are not independent (i.e.,
the possible SSS solutions are restricted to a discrete
set of constant energy surfaces within $\Lab$).
The fact that the consistency requirement on the Hamiltonian
formulation leads to the
`nonstandard' Hamiltonian framework for these cases
(i.e the foliation of phase space by
the  (symplectic) leaves on which there is a truly Hamiltonian framework)
together with the constancy of the full  Hamiltonian on stationary solutions
can be regarded as explaining such relation.
That is, on the intersection of each leave with the manifold ${\cal S}$
of stationary solutions, the full Hamiltonian is a
constant. In the limit $a_\Delta\mapsto 0$ on each
leave (provided such limit can be taken)
we will find a soliton.
The motion along the `leave' in ${\cal S}$ determines the mutual dependence
of the Isolated Horizon parameters.

\item %Near-horizon No-hair conjecture
It is now a general belief that the no-hair conjecture, even in
its weakest form \cite{mh,bizon2}, is violated for some
systems. One can thus hope that there be a
uniqueness result for static solutions
in terms of quasi-local parameters at the horizon.
In Section~\ref{sec6} we have put forward a `horizon parameters uniqueness
conjecture' stating that all static BH solutions are characterized by
their horizons parameters (`quasi-local charges') in a unique way.
In particular, it should be true that,
given the horizon area $a_\Delta$ and the
horizon electric charge $Q_\Delta$ and magnetic charge $P_\Delta$, the
Static solutions  be uniquely determined.
For instance, if we set $Q_\Delta$, given an arbitrary value
for  the magnetic charge $P_\Delta\in [0,1]$, there might be no
solution, but if there is a solution, it should be unique.
The numerical evidence available supports the conjecture
(see Figure~\ref{magnetic}).

\item One other point already mentioned is the issue of
the stability test
provided by this type of analysis: It is only when $M_{\rm ADM}>
M_\Delta$ in (\ref{7.5}) that the solution can be unstable.
One very clear
example of this is given by the magnetic RN solution,
which can be considered within both the Einstein Maxwell (EM)
theory and the EYM theory. This solution is stable within EM but
unstable within EYM \cite{bizon3,Kim,aich}. We can
now understand this situation
in the following way:  In the former case the gauge connection
can not be globally described through a gauge field $A_a$ as it
corresponds to a nontrivial bundle. In this case one can
nevertheless apply
the IH formalism  through the use of the duality symmetry of
Maxwell theory. This
results in the appearance of a term
$\Phi_M P_{\Delta}$ taking the place of
$\Phi Q_{\Delta}$ in (\ref{smarr}), and as  it is well known
the evaluation of
$E$ as $M_{\rm ADM} -M_{\Delta}$ (in this case $H^{(0)}=0$ as
there are no Abelian magnetically charged regular solitons)
gives $E=0$,
thus accounting for the stability of the solution. Let's look at what
happens in the EYM theory.
In this case the solution can be described in terms of the
gauge fields $A^i_a$ because it is associated with a trivial bundle
(with larger group) and there is therefore no term of the form
$\Phi_M P_{\Delta}$ in (\ref{smarr}). The $P_\Delta$ dependence
of the Mass $M_\Delta$ comes through $V$.
As we have shown, the value of $E$ for this solutions
is positive ($E=1$),
thus allowing for the instability of these solutions.
Note that this same argument is valid also for dyons with both
electric and unit magnetic charge (See Equation (\ref{6.24}) to
(\ref{6.27})), and thus, it indicates a potential instability
of these solutions.

Finally, let us conclude this remark by suggesting a `rule of thumb'
for finding potentially unstable solutions, suggested by the EYM
system. In the static family of solutions, consider the limit
$r_\Delta\mapsto 0$.
We have three possibilities: i) We arrive at a regular
solution with zero energy (i.e. Minkowski). This indicates that
the whole family, labeled by $r_\Delta$, is stable; ii) There
is a minimum allowed value of $r_\Delta$ corresponding to zero
surface gravity. In this case, we can not conclude anything, and;
iii) In the limit one finds a regular solution with positive energy
(a soliton different from the vacuum). In this case,
the whole family of solutions (including
 the soliton) is potentially unstable.
It would be interesting to re-examine,
from this perspective, the (complete non-linear)
stability of the Einstein-Skyrme Solitons and Black Holes \cite{skyrme}.

\end{enumerate}

\section{Discussion}
\label{sec9}

In this note, we have studied the extension  of the
Isolated Horizon formalism to include the EYM system and
found that it leads to a `nonstandard' Hamiltonian formulation.
The main feature of this formulation is that it provides a foliation of
phase space into symplectic leaves  in each of which we do get
a standard  Hamiltonian formulation.
The framework nevertheless  provides   a powerful
tool for studying some classical aspects of the theory
already at the Static level. In particular,
we found an  unexpected relation between the the ADM mass of
a static spherically symmetric
black hole solution, its Horizon mass and the ADM mass of the
corresponding solitonic
solution. These relationships were checked numerically in terms of the
known numerical results obtained in the process of finding
those solutions
and thus the agreement can be seen as a check on the whole formalism.

An apparent tension and a challenge for
the formalism is given by the existence of hairy
solutions, where the number of charges at infinity and at the horizon
do not coincide. We have been able to pin-point the problem
in a precise way using the Isolated Horizons formalism. This
involves the analysis of the
consistency of the Hamiltonian formulation and the nature of the
first law. We have encountered difficulties in defining a canonical
normalization for the vector $\l^a$ and thus, for the Horizon Mass 
in general.
We have proposed possible resolutions for this
difficulty.  Motivated by all these
results, and in order to have a  satisfactory treatment
of the EYM system within the framework,
we have put forward a `quasi-local uniqueness ' and a
`completeness' conjectures for Stationary Black holes.
In the case of the later we have put forward arguments both in favor and
against it,  but the main point is that some version of it
seems to be the only possibility to have the Isolated Horizon framework
working in EYM theory to the same extent  that it does in, say, 
Einstein-Maxwell theory.

The present work can be generalized in several directions. First,
our analysis allows us to propose a
Isolated Horizon treatment
for general theories containing non-trivial black holes and soliton
solutions; it should be possible to apply the
type of analysis presented here to these
theories where nontrivial regular
static solutions have been found. In particular,
Einstein-Yang-Mills-Higgs, Einstein-Yang-Mills-Dilaton,
and  Einstein-Skyrme Theories, are examples in
which there are both, solitonic  and
Black Hole solutions. In all these cases, formulas analogous to
the EYM case can in principle be found
by a straightforward application of the analysis carried out
in the last Section. In particular, one should be able to
to compute the `total energy' of the `hairy' black holes to test
for a potential instability.

Second,
one can use the very recent
results of Ashtekar and collaborators who have been able to
extend the isolated horizons framework to
include \textit{distorted} horizons \cite{afk} as well as
\textit{rotating} horizons \cite{abl}. Thus, one should be able to
use the formalism for distorted horizons in order to study
colored black holes which are static but not spherically symmetric
\cite{kk}. This analysis would be a first check for the
`quasi-local uniqueness conjecture $C1$' that we have proposed
\cite{cns2}.
The discussion of the preceding sections suggests that by
restricting our attention to non-distorted, non-rotating horizons
we were
lead to a consistent but `incomplete' formalism in the EYM system;
a complete treatment
(i.e., one providing a canonical choice of `normalization' for
all $\Lab$, based on
stationary solutions which are contained in the  formalism)
of EYM isolated horizons should be given within the context
of distorted and rotating Isolated Horizons \cite{afk,abl}.
It would be interesting to investigate this matter
once the articles \cite{afk,abl} are made public.

Finally, the present analysis shows that the so called colored black
hole solutions provide a nontrivial testing ground for the
approach of\cite{abck} to evaluate the statistical mechanical
entropy of black holes. In particular
it is known  that by selecting the value
$\gamma = \ln 2 / (\pi \sqrt 3 )$
for the Immirzi parameter $\gamma$,
(which amounts to selecting one  among a
continuous choice of unitarily inequivalent
quantum theories corresponding to the same classical theory),
the standard result $ S = A/4l_P^2 $ is obtained for the
Einstein Vacuum and
Einstein-Maxwell case. It is worth to
point out that this choice can be made only once, and that  it
is conceivable that say,
the choice needed in the case of Einstein-Maxwell black holes
might have been different than the choice needed for
pure gravity black holes. Now, it would be of
interest to check whether these results are also valid
when non-Abelian gauge fields are present.

\section*{Acknowledgments}

We would like to thank A. Ashtekar, C. Beetle, S. Fairhurst and
R. Wald for discussions and correspondence. We are also grateful
to the Center for Gravitational Physics and Geometry for its
hospitality. This work was in part supported by
DGAPA-UNAM grant No IN121298, by CONACyT
grants J32754-E and 32272-E, by a NSF-CONACyT collaborative
grant, by NSF grants INT9722514, PHY95-14240 and by the
Eberly research funds of Penn State.

\end{document}